\documentclass[aps,twocolumn,tightenlines,nofootinbib,superscriptaddress,floatfix]{revtex4-2}
\usepackage[utf8]{inputenc}
\usepackage{graphicx}
\usepackage{subcaption}
\usepackage{mwe}
\usepackage{mathrsfs}
\usepackage{amsfonts}
\usepackage{setspace}
\usepackage{cellspace}
\usepackage{amsmath,amssymb,bm}
\usepackage[colorlinks=true,linkcolor=blue]{hyperref}
\usepackage{xcolor}
\usepackage{epsfig}
\usepackage{slashed}
\usepackage{caption}
\usepackage{hhline,multirow,tabularx}  %
\usepackage{dcolumn}    %
\usepackage{url}        %
\usepackage{braket}     %
\usepackage{makecell}
\usepackage{bbm}
\usepackage{subcaption}

\newcommand{\dd}{\text{d}}

\DeclareMathAlphabet{\mathdutchcal}{U}{dutchcal}{m}{n}
\SetMathAlphabet{\mathdutchcal}{bold}{U}{dutchcal}{b}{n}
\DeclareMathAlphabet{\mathdutchbcal}{U}{dutchcal}{b}{n}

\begin{document}

\title{Energy-Energy Correlators in $e^+e^-$ and Deep Inelastic Scattering }
\author{Yuxun Guo}
\email{yuxunguo@lbl.gov}
\affiliation{Nuclear Science Division, Lawrence Berkeley National
Laboratory, Berkeley, CA 94720, USA}
\affiliation{Physics Department, University of California, Berkeley, California 94720, USA}

\author{Werner Vogelsang}%
 \email{werner.vogelsang@uni-tuebingen.de}
\affiliation{Institute for Theoretical Physics,
                Universit\"{a}t T\"{u}bingen,
                Auf der Morgenstelle 14,
                D-72076 T\"{u}bingen, Germany}

\author{Feng Yuan}
\email{fyuan@lbl.gov}
\affiliation{Nuclear Science Division, Lawrence Berkeley National
Laboratory, Berkeley, CA 94720, USA}

\author{Wenbin Zhao}
\email{wenbinzhao@ccnu.edu.cn}
\affiliation{Institute of Particle Physics and Key Laboratory of Quark and Lepton Physics (MOE), Central China Normal University, Wuhan, 430079, Hubei, China}
\affiliation{Nuclear Science Division, Lawrence Berkeley National
Laboratory, Berkeley, CA 94720, USA}

\begin{abstract}
We study energy-energy correlators (EECs) in $e^+e^-$ annihilation and deep inelastic lepton-hadron scattering (DIS), focusing on aspects of nonperturbative physics in these observables. We introduce the EEC jet functions and investigate the infrared (IR) behavior of both small-angle EECs and angle-integrated EECs by performing explicit one-loop calculations. The factorization and universality of the EECs in these processes are demonstrated. A matching scheme is proposed to smoothly connect kinematic regions where different scaling behaviors with jet energy are observed. In combination with the next-to-leading order correction, this matching provides a good description of the EEC data and PYTHIA simulations in high-energy $e^+e^-$ annihilation. Predictions for DIS processes for future electron-ion collider kinematics are also presented.
\end{abstract}
 
\maketitle

\section{Introduction}

Energy-energy correlators (EECs) in $e^+e^-$ annihilation and in jets in $pp$, $pA$ and $AA$ collisions have attracted great attention from both theory and 
experiment~\cite{Basham:1978bw,Basham:1977iq,Basham:1978zq,PLUTO:1985yzc,PLUTO:1979vfu,CELLO:1982rca,JADE:1984taa,Fernandez:1984db,Wood:1987uf,TASSO:1987mcs,AMY:1988yrv,TOPAZ:1989yod,ALEPH:1990vew,L3:1991qlf,L3:1992btq,DELPHI:1990sof,OPAL:1990reb,OPAL:1991uui,SLD:1994idb,Collins:1981uk,Ali:1982ub,Clay:1995sd,deFlorian:2004mp,DelDuca:2016csb,Tulipant:2017ybb,Kardos:2018kqj,Moult:2018jzp,Dixon:2018qgp,Dixon:2019uzg,Ebert:2020sfi,Schindler:2023cww,Lee:2006nr,Berger:2003iw,Hofman:2008ar,Chen:2020vvp,Lee:2022ige,Craft:2022kdo,Komiske:2022enw,Andres:2022ovj,Andres:2023xwr,Andres:2023ymw,Yang:2023dwc,Andres:2024ksi,Barata:2023bhh,Barata:2023zqg,Bossi:2024qho,Lee:2023tkr,Lee:2023xzv,Lee:2024esz,Chen:2024nyc,Holguin:2023bjf,Holguin:2024tkz,Xiao:2024rol,Xing:2024yrb,Andres:2024hdd,Liu:2024lxy,Alipour-fard:2024szj,Kang:2024dja,Barata:2024wsu,Csaki:2024zig,Fu:2024pic,Apolinario:2025vtx,Barata:2025fzd,Chen:2025rjc,Moult:2025nhu,CMS:2024mlf,ALICE:2024dfl,Tamis:2023guc,CMS:2025ydi,ALICE:2025igw,ALICEpA,Barata:2024ukm,Lee:2025okn,Chang:2025kgq,Guo:2025zwb,Herrmann:2025fqy,Kang:2025zto,Zhao:2025ogc,Jaarsma:2025tck}. They provide unique opportunities to study features of strong-interaction physics and quantum chromodynamics (QCD). In the past few years, significant progress has been made on all frontiers~\cite{Moult:2025nhu}.

Formally, the EEC is defined as
\begin{equation}
\frac{1}{\sigma_{\text{tot}}} \frac{\dd\Sigma}{\dd\cos\theta} = \frac{1}{\sigma_{\text{tot}}} \sum_{i\neq j} \int \dd\sigma^{ij} \, \frac{E_i E_j}{Q^2} \, \delta(\cos\theta - \hat{n}_i \cdot \hat{n}_j), \label{eq:eec0}
\end{equation}
where $\dd\sigma^{ij}$ denotes the semi-inclusive differential cross-section for producing two hadrons $h_i$ and $h_j$, with energies $E_i$ and $E_j$, and in the direction of the unit vectors $\hat{n}_i$ and $\hat{n}_j$, respectively. The total energy is $Q = \sum_i E_i$, and $\theta$ is the angle between the two directions. For the $e^+e^-$ annihilation process, one further defines for convenience the variable $\zeta \equiv (1 - \cos\theta)/2$. Studies of EECs have focused on 
two distinct kinematic regions: the ``near side'' with $\zeta\to 0$, and the ``away side'' where $\zeta\to 1$. In the 
latter regime, the hadrons are produced ``back-to-back'', and their 
total transverse momentum $q_T$ is much smaller than the total energy $Q$. The EEC is then dominated by soft and collinear gluon radiation and an appropriate transverse-momentum dependent (TMD) factorization has been applied~\cite{Collins:1981uk,Collins:1981uw,Collins:1981va,Collins:1981zc}, which incorporates an all-order resummation
of double logarithms in $q_T/Q$~\cite{Collins:1984kg,deFlorian:2004mp}. The TMD resummation has been included in phenomenological studies~\cite{deFlorian:2004mp,Tulipant:2017ybb,Kardos:2018kqj,Ebert:2020sfi,Kang:2024dja} of the EEC measurements in $e^+e^-$ annihilation. 

In this paper, we study the hadronization effects in the EEC observables, focusing on the nonperturbative aspects of the measurements. Previously, QCD factorization for the EECs has been established in the collinear limit, and a cumulant method was applied~\cite{Dixon:2019uzg}. Our analysis follows the direction given in that reference. In particular, by carefully examining the infrared (IR) behavior of Eq.~(\ref{eq:eec0}), one realizes that although for any finite $\zeta$ the EEC is IR safe, upon integration over $\zeta$, IR safety is lost. The angular-integrated EEC actually depends on the di-hadron fragmentation functions introduced in the literature in other contexts for hadron physics studies~\cite{Konishi:1978yx,Konishi:1979cb,Vendramin:1981te,deFlorian:2003cg,Majumder:2004wh,Majumder:2004br,Collins:1993kq,Ji:1993vw,Jaffe:1997hf,Bianconi:1999cd,Bacchetta:2002ux,Bacchetta:2012ty,Zhou:2011ba,Cocuzza:2023vqs,Pitonyak:2023gjx,Rogers:2024nhb}. 

To describe the collinear behavior for the EEC observables, we introduce EEC jet functions, which can be constructed from di-hadron fragmentation functions, summing over all hadrons weighted with their energy fractions. We will also show that the jet functions are universal across different processes. The integrated EEC jet functions only depend on a factorization scale. This scale dependence can be simply obtained from the evolution of specific moments of the fragmentation functions. Interestingly, because of momentum conservation in the di-hadron fragmentation functions~\cite{Konishi:1979cb,deFlorian:2003cg}, the EEC jet functions have a very mild scale dependence as we will demonstrate below. This may give EEC observables an advantage as tools for studying hadronic structure. 

We further extend these ideas to the unintegrated EEC jet functions depending on the opening angle (or, transverse momentum difference) 
between the two final-state hadrons. We will perform one-loop calculations to investigate the IR behavior at small angles, and derive an evolution equation in the Fourier transform $b_T$-space similar to that of Ref.~\cite{Dixon:2019uzg}. Following the strategy of the TMD approach in~\cite{Collins:1984kg,Boussarie:2023izj}, the $b_T$-space allows us to build connections between the integrated EEC jet functions and the unintegrated ones and carry out an all-order resummation of transverse-momentum logarithms. Different from the usual TMD resummation, for the unintegrated EEC jet function, there are no Sudakov double logarithmic contributions, so that we only need to resum single logarithms. 

To demonstrate the universality of the unintegrated EEC jet functions we carry out one-loop calculations of the EEC observables in $e^+e^-$ annihilation and DIS. We show the factorization of the near-side EECs in terms of these unintegrated EEC jet functions. With EEC jet functions determined from $e^+e^-$ annihilation, we present predictions for DIS for typical kinematics at the future electron-ion collider (EIC). 

Several other approaches have been proposed recently for applying the di-hadron fragmentation function formalism to the near-side EECs~\cite{Lee:2025okn,Kang:2025zto}. Different from these papers, we focus on the IR behavior of the EEC jet function and derive the factorization and resummation formalism based on explicit one-loop calculations.
A brief summary of our results has been reported in Ref.~\cite{Guo:2025zwb}. 

The rest of the paper is organized as follows. In Sec.~II, we introduce the EEC jet functions. We formulate them in terms of general di-hadron fragmentation functions, summing over all final-state hadrons and weighting with their energy fractions. We also explore the behavior of the EEC jet functions at small angles. In Sec.~III, we investigate the EECs in $e^+e^-$ annihilation. We first study the integrated EECs in terms of the jet functions introduced in the preceding section. Comparison of our results to PYTHIA event simulations supports the picture of collinear evolution of the EEC jet functions. We then focus on the near-side small-angle EECs in $e^+e^-$, where we demonstrate 
factorization and derive resummation. Section~IV is devoted to the EECs in DIS lepton-nucleon collisions. We carry out the relevant next-to-leading order calculations. For the near-side small-angle EECs, we apply resummation in terms of the unintegrated EEC jet functions. 

\section{EEC Jet Function}

For any generic process that produces two ``observed'' hadrons in the final state, we have the following two types of contributions~\cite{deFlorian:2003cg}:
\begin{eqnarray}
    \frac{\dd\sigma^{(h_1h_2)}}{\dd z_1\dd z_2}&=&\sum_{ij} \int\frac{\dd  x_1}{x_1}\frac{\dd x_2}{x_2}\frac{\dd\hat\sigma_{ij}}{\dd x_1\dd x_2}D_i^{h_1}\left(\frac{z_1}{x_1}\right)D_j^{h_2}\left(\frac{z_2}{x_2}\right)\nonumber\\
    &&+\sum_i\int\frac{\dd x}{x^2}\frac{\dd \hat \sigma_i}{\dd x} I\!\!D_i^{h_1h_2}\left(\frac{z_1}{x},\frac{z_2}{x}\right) \ ,\label{eq:dihadron}
\end{eqnarray}
where the first term represents two individual fragmentation processes, for which two partons $i$ and $j$, produced via a hard-scattering 
cross section $\hat\sigma_{ij}$, give rise to the two hadrons through standard single-hadron fragmentation functions $D_i^h$. For the second term in~(\ref{eq:dihadron})
the pair of hadrons comes from a single parton $i$ that is produced with partonic cross section $\hat\sigma_i$ and fragments 
according to a di-hadron fragmentation function $I\!\!D_i^{h_1h_2}$. In the above equation, $x_{1,2}$ and $x$ represent the energy (momentum) fractions 
associated with the fragmenting partons. For simplicity, we omit all other kinematic variables and the scale dependence in the fragmentation functions. 

If we sum over all hadrons in the final state, we can immediately derive sum rules associated with 
momentum conservation in the fragmentation process~\cite{Konishi:1979cb,deFlorian:2003cg},
\begin{eqnarray}
&&    \int \dd z z \sum_hD_i^h(z,\mu)=1 \label{eq:sumrule1}\,, \\
&&    \int \dd z_1 z_1\sum_{h_1}I\!\!D_i^{h_1h_2}(z_1,z_2;\mu)=(1-z_2)D_i^{h_2}(z_2,\mu) \, . \;\;\;\;\label{eq:sumrule2}
\end{eqnarray}
In particular, the second equation builds a connection between the di-hadron fragmentation functions and the single-hadron ones. The above equations become an important tool when we study the EEC observables. 

\subsection{Integrated EEC Jet Function\label{sec:int}}

Applying the above idea to the EEC observables, we should also have two contributions. Therefore, a generic EEC observable can be computed from the following di-hadron differential cross section,
\begin{eqnarray}
   && \int \dd z_1 \dd z_2 z_1z_2\frac{\dd \sigma(h_1h_2)}{\dd z_1\dd z_2}\nonumber\\
    &&~~~~~=\int \dd z_1 \dd z_2 z_1 z_2\sum_{ij,h_1h_2}\hat \sigma_{ij}D_i^{h_1}(z_1)D_j^{h_2}(z_2)\nonumber\\
    &&~~~~~+\int \dd z_1 \dd z_2 z_1 z_2\sum_{i,h_1h_2}\hat \sigma_i I\!\!D_i^{h_1h_2}(z_1,z_2) \ . \label{eq:eec_general}
\end{eqnarray}
For the EEC observables, we sum over all final-state hadrons. Clearly, the first term can be simplified with just partonic calculations because of the momentum conservation 
relation (\ref{eq:sumrule1}), and it can be computed in perturbative QCD. However, the second term will inevitably involve nonperturbative physics because of
the residual fragmentation function. To describe this physics, we introduce an integrated EEC jet function:
\begin{equation}
    \Gamma_i(\mu)\equiv\sum_{h_1h_2}\int \dd z_1\dd z_2 z_1 z_2 I\!\!D_i^{h_1h_2}(z_1,z_2,\mu) \ .\label{eq:dihadron0}
\end{equation}
According to the momentum sum rule (\ref{eq:sumrule2}), the above $\Gamma_i(\mu)$ can also be simplified as
\begin{equation}
    \Gamma_i(\mu)=1-\Gamma_i'(\mu) \ , \label{eq:gamma0}
\end{equation}
where $\Gamma_i'(\mu)$ is the third moment of the single-hadron fragmentation function,
\begin{equation}
    \Gamma_i'(\mu)=\sum_h\int \dd z z^2 D_i^h(z,\mu)\ .
\end{equation}
We are now ready to derive the evolution equations for the above nonperturbative quantities. These are necessarily related to the (time-like) DGLAP evolution of the fragmentation functions, which are well known to take the following form:
\begin{equation}
    \frac{\dd}{\dd \log \mu^2} \boldsymbol{D}^H(z, \mu) = \left(\boldsymbol{D}^H\otimes \hat{{\boldsymbol{P}}}_T\right)(z,\mu) \ .
\end{equation}
Here we have written the fragmentation functions as flavor-space vectors: $\boldsymbol{D}^H(z, \mu^2)\equiv(D_q^H,D_g^H)$, where the singlet quark combination 
is defined as $q=\sum_{f}(q_f+\bar q_f)$. The time-like DGLAP splitting kernel $\hat{{\boldsymbol{P}}}_T$ is then a matrix in flavor space:
\begin{equation}
    \hat{{\boldsymbol{P}}}=\begin{pmatrix}
    P_{qq} & P_{q g}\\
     P_{gq} & P_{gg}
    \end{pmatrix}\ ,
\end{equation}
where we dropped the $T$ subscript for simplicity and implicitly assume a time-like DGLAP kernel hereafter. In Mellin moment space, the splitting kernel $\hat{{\boldsymbol{P}}}$ is also known as the anomalous dimension matrix,
\begin{equation}
    \hat{\boldsymbol{\gamma}}^{(N)}\equiv- \int \dd x \ x^{N-1} \hat{\boldsymbol{P}}(x)\ .
\end{equation}
We define the Mellin moments of the fragmentation functions accordingly as:
\begin{equation}
    \overline{\boldsymbol{D}}^H(N; \mu) \equiv \int\dd z\  z^{N-1} \boldsymbol{D}^H(z; \mu) \ ,
\end{equation}
whose evolution equations read:
\begin{equation}
    \frac{\dd}{\dd \log \mu^2} \overline{\boldsymbol{D}}^H(N; \mu) = -\overline{\boldsymbol{D}}^H(N;\mu)\cdot\hat{{\boldsymbol{\gamma}}}^{(N)} \ ,
\end{equation}
according to the Mellin convolution theorem. From the relation $\Gamma'(\mu)=\overline{\boldsymbol{D}}^H(N=3; \mu)$ it is now obvious that 
\begin{equation}
\begin{split}
\frac{\dd}{\dd \log{\mu^2}} \boldsymbol\Gamma'(\mu)
    =-\boldsymbol\Gamma'(\mu)\cdot \hat{{\boldsymbol{\gamma}}}^{(N=3)} \ .
\end{split}
\end{equation}
Combining the above with Eq.~(\ref{eq:gamma0}), we find the evolution of $\Gamma_i(\mu)$, which contains an inhomogeneous term:
\begin{equation}
\frac{\dd}{\dd \log{\mu^2}} \boldsymbol\Gamma(\mu)=-\boldsymbol\Gamma(\mu)\cdot\hat{{\boldsymbol{\gamma}}}^{(3)} +\boldsymbol{1}\cdot\hat{{\boldsymbol{\gamma}}}^{(3)}\ ,
\end{equation}
where the second term is the inhomogeneous contribution. The above evolution can be verified by applying the scale evolution for di-hadron fragmentation functions~\cite{deFlorian:2003cg} in Eq.~(\ref{eq:dihadron0}). In practice, we can solve the evolution equation for $\Gamma'$ and make predictions for the integrated EEC.

In what follows, we perturbatively expand the anomalous dimensions and splitting kernels as
\begin{eqnarray}
    \hat{\boldsymbol{\gamma}}^{(N)} &=& \sum_{n} \left(\frac{\alpha_s}{2\pi}\right)^{\,n+1} \hat{\boldsymbol{\gamma}}^{(N)}_{n},\nonumber\\
    \hat{\boldsymbol{P}}(x) 
    &=& \sum_{n} \left(\frac{\alpha_s}{2\pi}\right)^{\,n+1} \hat{\boldsymbol{P}}_{n}(x),
\end{eqnarray}
where $\alpha_s$ is the strong coupling. For simplicity of notation, we refer to the leading-order terms 
$\hat{\boldsymbol{\gamma}}^{(N)}_{0}$ and $\hat{\boldsymbol{P}}_{0}(x)$ simply again as 
$\hat{\boldsymbol{\gamma}}^{(N)}$ and $\hat{\boldsymbol{P}}(x)$ throughout this paper.

The leading-order anomalous dimensions then read:
\begin{align}
    \gamma_{qq}^{(N)} &= -\frac{C_F}{2} \left[3+\frac{2}{N(N+1)}-4S_1(N) \right]\,,\\
    \gamma_{gq}^{(N)} &=-  C_F \frac{N(N+1)+2}{(N-1)N(N+1)}\,,\\
    \gamma_{qg}^{(N)} &= -2T_F n_f \frac{N(N+1)+2}{N(N+1)(N+2)}\,,\\
    \begin{split}
    \gamma_{gg}^{(N)} &= \frac{C_A}{2} \Bigg[\frac{4}{N(N+1)}-\frac{12}{(N-1)(N+2)}\\
    &\qquad\qquad+4S_1(N)\Bigg]-(11 C_A -4 T_F n_f)/6\,,
    \end{split}
\end{align}
where $C_F=4/3$, $C_A=3$, $T_F=1/2$ and $n_f$ is the number of active flavors. The harmonic sum $S_1(N)\equiv \psi(N+1)+\gamma_E$ can be expressed in terms of the digamma function $\psi$ and the Euler constant $\gamma_E$. Specifically for $N=3$, one has
\begin{equation}
    \hat{\boldsymbol{\gamma}}^{(3)}=\begin{pmatrix}
    \frac{25}{12}C_F & -\frac{7n_f}{30}\\[1mm]
    -\frac{7}{12}C_F & \frac{7}{5}C_A+\frac{1}{3}n_f
    \end{pmatrix}\ .\label{eq:gamma3t}
\end{equation}
This matrix has two positive eigenvalues, approximately 6.14 and 2.51 for $N_c=3$ and $n_f=5$, indicating that asymptotically $\Gamma_{q,g}'(\mu)\to0$ and therefore $\Gamma_{q,g}(\mu)=1-\Gamma_{q,g}'(\mu)\to1$ in the $\mu\to\infty$ limit.

\subsection{Unintegrated EEC Jet Function \label{sec:unintEEC}}

We now go one step further and turn to the angular distribution of the EEC. Similar to the integrated EEC discussed above, in the near-side region at small angle, the EEC also depends on nonperturbative physics~\cite{Dixon:2019uzg,Liu:2024lxy}. To study the IR behavior, we introduce the unintegrated EEC jet function, again in terms of a general di-hadron fragmentation function:
\begin{eqnarray}
    \Gamma_i(\mu,q_T)&=&\sum_{h_1h_2}\int \dd ^2p_{1T}\dd ^2p_{2T} \dd z_1 \dd  z_2 (z_1 z_2) \nonumber\\
    &&\times I\!\! D_i^{h_1h_2}(z_1,z_2,p_{1T},p_{2T})\nonumber\\
    &&\times \delta^{(2)}\left(\vec{q}_T-\left(\frac{\vec p_{1T}}{z_1}-\frac{\vec p_{2T}}{z_2}\right)\right) \ ,\label{eq:ddqt}
\end{eqnarray}
where $p_{1T}$ and $p_{2T}$ are the transverse momenta of the two final-state hadrons with respect to the fragmenting quark or gluon $i$ %
and $z_{1,2}$ are the longitudinal momentum fractions. In the following, we will study the IR behavior for the unintegrated EEC jet function and derive an all-order resummation based on factorization. 

The key element of our treatment is the Fourier transform to $b_T$-space,
\begin{equation}
    \widetilde{\Gamma}_q(\mu,b_T)=\int {\dd^2q_T} \ e^{i\vec{q}_T\cdot \vec{b}_T}\Gamma_q(\mu,q_T) \ ,
\end{equation}
for the quark EEC jet function as an example. There is also scale dependence on both sides of the equation. Applying Eq.~(\ref{eq:ddqt}), we can also write
\begin{eqnarray}
    \widetilde{\Gamma}_i(\mu,b_T)&=&\sum_{h_1h_2}\int {\dd^2p_{1T}\dd ^2p_{2T}} \dd z_1 \dd z_2 (z_1 z_2) \label{eq:ddbt}\\
    &&\times e^{i \vec{b}_T\cdot\left(\frac{\vec{p}_{1T}}{z_1}-\frac{\vec{p}_{2T}}{z_2}\right)}I\!\! D_i^{h_1h_2}(z_1,z_2,p_{1T},p_{2T})\nonumber \ .
\end{eqnarray}
To illustrate the factorization argument, let us again go to perturbation theory. The leading-order result for $\widetilde{\Gamma}$ is 
\begin{equation}
    \widetilde{\Gamma}^{(0)}_q(\mu,b_T)=\int\frac{\dd^2q_T}{(2\pi)^2} e^{i\vec{q}_T\cdot \vec{b}_T}\Gamma^{(0)}_q(\mu,q_T) \ ,
\end{equation}
where we only assume that the nonperturbative jet function peaks at low $q_T$. At one-loop order, we have homogeneous and inhomogeneous contributions, similar to what we discussed in the previous subsection. 

\begin{figure}[htbp]
 \begin{center}
   \includegraphics[width=0.45\textwidth]{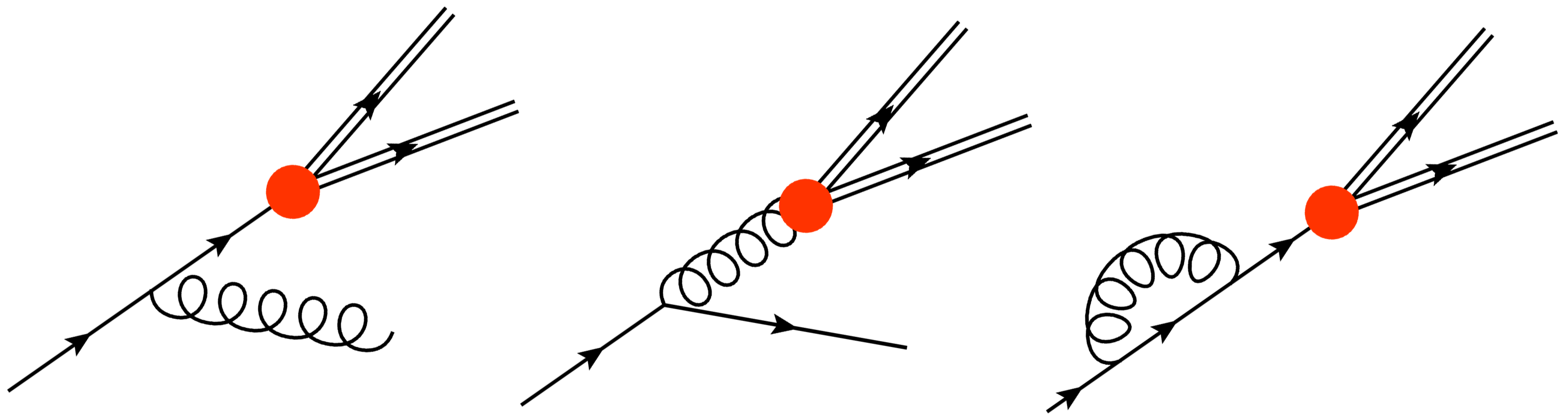} 
\caption{
Homogeneous contributions at one-loop order. The transverse momentum difference between the hadrons 
with respect to the initial quark's momentum, $q_T=|\left(\frac{\vec p_{1T}}{z_1}-\frac{\vec p_{2T}}{z_2}\right)|$,
remains unchanged. However, with respect to the daughter quark or gluon, $q_T'=xq_T$, where $x$ is the momentum fraction carried by the daughter parton.}
  \label{fig:eec_spliting}
\end{center}
  \vspace{-5.ex}
\end{figure}

An important feature for the homogeneous contributions is that they do not change the momentum difference $q_T=|\left(\frac{\vec p_{1T}}{z_1}-\frac{\vec p_{2T}}{z_2}\right)|$
(see Fig.~\ref{fig:eec_spliting}). However, the longitudinal momentum distribution will be modified via the usual DGLAP splitting kernel. For example, from the $q\to q(+g)$ splitting contribution, we have
\begin{equation}
    \left.{\Gamma}^{(1)}_q
    \right|_{\mathrm{hom.}}=
    \frac{\alpha_s}{2\pi}\left(-\frac{1}{\hat \epsilon}+\ln\frac{\mu^2}{\bar\mu^2}\right)\int {\dd x}{x^4} \hat P_{qq}(x) \Gamma_q^{(0)}(\mu,q_T') \ , 
\end{equation}
where $\mu=E$ represents the scale for the EEC jet function and $q_T'=xq_T$ denotes the transverse momentum difference between the two hadrons relative to the daughter quark.
This translates to the following result in Fourier transform $b_T$-space:
\begin{equation}
    \left.\widetilde{\Gamma}^{(1)}_q
    \right|_{\mathrm{hom.}}= 
    \frac{\alpha_s}{2\pi}\left(-\frac{1}{\epsilon}+\ln\frac{\mu^2}{\bar\mu^2}\right)
    \int{\dd x}{x^2} \hat P_{qq}(x)\widetilde{\Gamma}_q^{(0)}\left(\mu,{b_T\over x}\right)  \ . 
\end{equation}
A similar result can be derived for the $q\to g(+q)$ channel. The total contribution from the homogeneous terms can be written as
\begin{eqnarray}
&&    \left.\widetilde{\Gamma}^{(1)}_q(\mu,b_T)
    \right|_{\mathrm{hom.}}\nonumber\\
    &&~~~=
    \frac{\alpha_s}{2\pi}\left(-\frac{1}{\epsilon}+\ln\frac{\mu^2}{\bar\mu^2}\right)
    \int{\dd x}{x^2} \left\{\hat P_{qq}(x) \widetilde{\Gamma}_q^{(0)}\left(\mu,{b_T\over x}\right) \right.\nonumber\\
    &&~~~~~~~~~~~~~\left.+\hat P_{gq}(x) \widetilde{\Gamma}_g^{(0)}\left(\mu,{b_T\over x}\right) \right\}\ . 
\end{eqnarray}

\begin{figure}[htbp]
 \begin{center}
   \includegraphics[width=0.18\textwidth]{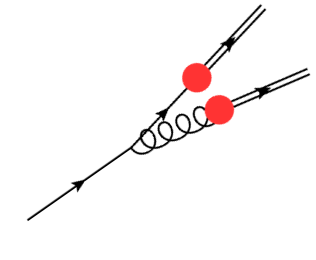} 
\caption{
Inhomogeneous contributions at one-loop order. }
  \label{fig:eec_spliting1}
\end{center}
  \vspace{-5.ex}
\end{figure}

The inhomogeneous contributions are represented by the diagrams in Fig.~\ref{fig:eec_spliting1}. By Fourier transforming the $q_T$ dependence to $b_T$-space, we have
\begin{equation}
    \left.\widetilde{\Gamma}^{(1)}_q(\mu,b_T)
    \right|_{\mathrm{inhom.}}=
    \frac{\alpha_s}{2\pi}\left(-\frac{1}{\epsilon}+\ln\frac{\mu_b^2}{\bar\mu^2}\right)\left(\gamma_{qq}^{(3)}+\gamma_{gq}^{(3)}\right)\ ,\label{eq:inhomobt}
\end{equation}
where $\mu_b=2e^{-\gamma_E}/b_T$. We have neglected a constant term in the above equation, which is sub-leading at the order we are working here. Adding both contributions, we have 
\begin{eqnarray}
    \widetilde{\Gamma}^{(1)}_q(\mu,b_T)&=&\frac{\alpha_s}{2\pi} \left(-\frac{1}{\epsilon}+\ln\frac{\mu^2}{\bar\mu^2}\right)
    \int{dx}{x^2} \left\{\hat P_{qq}(x) \right.\nonumber\\
    &&~~\left.\times \widetilde{\Gamma}_q^{(0)}\left(\mu,{b_T\over x}\right) +\hat P_{gq}(x) \widetilde{\Gamma}_g^{(0)}\left(\mu,{b_T\over x}\right) \right\}\nonumber\\
        &+& \frac{\alpha_s}{2\pi}\left(-\frac{1}{\epsilon}+\ln\frac{\mu_b^2}{\bar\mu^2}\right)\left(\gamma_{qq}^{(3)}+\gamma_{gq}^{(3)}\right) \ .
        \label{eq:oneloopbt0}
\end{eqnarray}
The IR divergence should be absorbed into the renormalization of the EEC jet function. After subtracting the pole, we can write, up to one-loop order,
\begin{eqnarray}
    \widetilde{\Gamma}_q(\mu,b_T)&=& \widetilde{\Gamma}_q(\mu_b,b_T)\nonumber\\
    &+&\frac{\alpha_s}{2\pi}\ln\frac{\mu^2}{\mu_b^2}
    \int{dx}{x^2} \left\{\hat P_{qq}(x) \widetilde{\Gamma}_q(\mu,{b_T\over x}) \right.\nonumber\\
    &&~~~~~~~~~~~~~\left.+\hat P_{gq}(x) \widetilde{\Gamma}_g(\mu,{b_T\over x}) \right\}\ , 
    \label{eq:oneloopbt1}
\end{eqnarray}
where the EEC jet functions are the renormalized ones.  From this we can derive an evolution equation for the unintegrated EEC jet function in $b_T$-space~\cite{Dixon:2019uzg,Barata:2024wsu,Herrmann:2025fqy}:
\begin{eqnarray}
    \frac{\partial}{\partial\ln\mu^2}\widetilde{\Gamma}_q(\mu,b_T)&=&\frac{\alpha_s}{2\pi}\int{dx} x^2\left\{\hat P_{qq}(x) \widetilde{\Gamma}_q\left(\mu,{b_T\over x}\right) \right.\nonumber\\
    &&~~~~~~\left.+\hat P_{gq}(x) \widetilde{\Gamma}_g\left(\mu,{b_T\over x}\right) \right\}\ .\label{eq:evolutionbt0}
\end{eqnarray}
We emphasize that the first term of Eq.~(\ref{eq:oneloopbt1}) provides the initial condition for the evolution equation.

Two important features can be derived from the above results. First, if we are in the perturbative region, i.e., $1/b_T\gg \Lambda_{\rm QCD}$, the scaling behavior of the EEC leads to a logarithmic dependence on $b_T$. Therefore, we can make a leading logarithmic approximation (LLA), and Eq.~(\ref{eq:oneloopbt1}) can be simplified to
\begin{eqnarray}
&&  \hspace*{-4mm}  \left.\widetilde{\Gamma}_q(\mu,b_T)\right|_{{\rm LLA}}= \widetilde{\Gamma}_q(\mu_b,b_T)\nonumber\\
&& \hspace*{-4mm}
~~~~~~~+\frac{\alpha_s}{2\pi}\ln\frac{\mu^2}{\mu_b^2}\left[\widetilde{\Gamma}_q(\mu,b_T)\gamma_{qq}^{(3)}+\widetilde{\Gamma}_g(\mu,b_T)\gamma_{gq}^{(3)}\right]\ ,\;\;\;
    \label{eq:oneloopscaling}
\end{eqnarray}
and the evolution equation becomes
\begin{equation}
      \left.\frac{\partial \widetilde{\Gamma}_q(\mu,b_T)}{\partial\ln\mu^2}\right|_{{\rm LLA}}=\frac{\alpha_s}{2\pi}\left(\widetilde{\Gamma}_q(\mu,b_T)\gamma_{qq}^{(3)}+\widetilde{\Gamma}_g(\mu,b_T)\gamma_{gq}^{(3)}\right) \ .\label{eq:evolutionbt}
\end{equation}
From this we can derive an all-order resummation~\cite{Guo:2025zwb}, for which we have also demonstrated that it reproduces the asymptotic expansion of the fixed-order calculations. 

On the other hand, in the nonperturbative region, i.e., $b_T\gg 1/\Lambda_{\rm QCD}$, it has been realized that the scaling of the EEC has a different behavior~\cite{Chang:2025kgq}. We will show in the following that this observation is consistent with the above one-loop result and the evolution equation. In this regime, to illustrate the scaling behavior in the infrared region, we can Fourier transform the evolution equation back to momentum space,
\begin{eqnarray}
    \frac{\partial {\Gamma}_q(\mu,q_T)}{\partial\ln\mu^2}&=&\frac{\alpha_s}{2\pi}\int{dx} x^4\left\{\hat P_{qq}(x){\Gamma}_q(\mu,xq_T) \right.\nonumber\\
    &&~~~~~~\left.+\hat P_{gq}(x){\Gamma}_g(\mu,x{q_T}) \right\}\ . \label{eq:evolutionqt}
\end{eqnarray}
Taking the small-$q_T$ limit, we arrive at
\begin{eqnarray}
    \frac{\partial {\Gamma}_q(\mu,0)}{\partial\ln\mu^2}&=&\frac{\alpha_s}{2\pi}\int{dx} x^4\left\{\hat P_{qq}(x){\Gamma}_q(\mu,0) \right.\nonumber\\
    &&~~~~~~\left.+\hat P_{gq}(x){\Gamma}_g(\mu,0) \right\}\ .
\end{eqnarray} 
This leads to the renormalization group equation
\begin{eqnarray}
    \frac{\partial {\Gamma}_q(\mu,0)}{\partial\ln\mu^2}&=&\frac{\alpha_s}{2\pi}\left({\Gamma}_q(\mu,0)\gamma_{qq}^{(5)}+{\Gamma}_g(\mu,0)\gamma_{gq}^{(5)}\right) \ ,\;\;\;
\end{eqnarray} 
which is consistent with the result in Ref.~\cite{Chang:2025kgq}. Furthermore, we can also expand Eq.~(\ref{eq:evolutionqt}) at small $q_T$ in terms of powers $(q_T^2)^n$. We note that in the Fourier inverse of the evolution equation~(\ref{eq:evolutionbt0}) from $b_T$- to $q_T$-space odd powers of $q_T$ vanish. 
A scaling behavior follows for each contribution with $(q_T^2)^n$ to the expansion of $\Gamma_q(\mu,q_T)$, with an associated 
anomalous dimension $\gamma_{ij}^{(5+2n)}$. This is again consistent with Ref.~\cite{Chang:2025kgq}.

It is important to note that in the perturbative region, i.e., $q_T\gg\Lambda_{\rm QCD}$, the unintegrated EEC jet function has a power behavior $\Gamma_q(\mu,q_T)\sim 1/q_T^2$. Substituting this into Eq.~(\ref{eq:evolutionqt}), we arrive at the same conclusion as that in Eq.~(\ref{eq:evolutionbt}), i.e., the anomalous dimension is $\gamma_q^{(3)}$ for large $q_T$, instead of the $\gamma^{(5)}$ for $q_T\to 0$.

In principle, one should be able to solve the above evolution equation with a certain initial condition, e.g., $ \tilde{\Gamma}_i(\mu=\mu_b,b_T)\approx \tilde{\Gamma}_i(\mu_b)\tilde{\Gamma}_i^{\rm NP}(b_T)$, where $\tilde{\Gamma}_i(\mu_b)$ is the integrated EEC jet function at scale $\mu=\mu_b$ and $\widetilde{\Gamma}_i^{\rm NP}(b_T)$ a nonperturbative model in $b_T$-space. This strategy follows the common procedure in TMD phenomenology~\cite{Boussarie:2023izj}. We expect that one can explore a number of methods to solve the evolution equation, especially an iterative method as used in Ref.~\cite{Herrmann:2025fqy}. The solution has to respect both constraints discussed above. We will come back to this in a future publication. In the following, we will make use of the above analytic insights to model the EEC jet functions with all-order resummation included.

\subsection{Resummation and Matching Between Different Scaling Regions \label{sec:match}}

Clearly, the EEC jet functions have different scaling behaviors in the perturbative and nonperturbative regions. This indicates it may not be possible to obtain a universal solution for their evolution equations, and a matching between the perturbative and nonperturbative regions will be necessary. A numerical solution is still possible for the evolution equations. In the following, we will discuss how to improve the LLA solution with additional constraints from the scaling in the nonperturbative region by performing such a matching. 

The LLA solution for the evolution takes the following form:
\begin{eqnarray}
&&\widetilde{\Gamma}_i^{\rm LLA}(\mu,b_T)\nonumber\\
&&=\Gamma_j(\mu_b) \text{P}{\exp
\left[-\int_{\mu_b^2}^{\mu^2}\frac{d\mu^{\prime 2}}{\mu^{\prime 2}}\frac{\alpha_s(\mu')}{2\pi} \gamma^{(3)}(\mu')\right]_{ji}} \ ,     
\end{eqnarray}
where $\text{P}$ denotes path ordering, and we parameterize the EEC jet function at the lower scale $\mu_b$
by the integrated EEC jet function $\Gamma_j$ of Sec.~\ref{sec:int}, also at scale $\mu_b$. When Fourier transformed back to $q_T$-space, the above resummation formula will encounter the nonperturbative region. We follow the so-called $b_*$ prescription~\cite{Collins:1984kg}, $b_T\to b_*=b_T/\sqrt{1+b_T^2/b_{\textrm{max}}^2}$, choosing $b_{\textrm{max}}=1.5\ \rm GeV^{-1}$. With that, we also include a nonperturbative contribution. Our final resummation result for $\widetilde{\Gamma}$ reads
\begin{equation}
    \widetilde{\Gamma}_{i,\textrm{res}}^{\rm LLA} (\mu,b_T)=\widetilde{\Gamma}_i^{\rm LLA}(\mu,b_*)e^{-\Lambda_i b_T} \ ,\label{eq:nonpert}
\end{equation}
where the $\Lambda_i$ ($i=q,g$) are free parameters that are fitted to experimental data. Following Refs.~\cite{Barata:2024wsu,Herrmann:2025fqy}, we have adopted a simple exponential form for the nonperturbative part of the EEC jet functions. With the above result, we obtain the LLA unintegrated EEC jet function in transverse-momentum space,
\begin{equation}
    \Gamma_i^{\rm LLA}(\mu,q_T)=\int\frac{d^2b_T}{(2\pi)^2}e^{iq_T\cdot b_T}\widetilde{\Gamma}_{i,\textrm{res}}^{\rm LLA} (\mu,b_T) \ .
\end{equation}
However, as mentioned above, at low transverse momentum the scaling behavior changes. 
In order to improve the prediction, we introduce a modified hard factor, defined by
\begin{equation}
\begin{split}
        {\cal C}_{ji}(\mu)=\sum_k&\,\text{P}\exp\left[\int_{\mu_0^2}^{\mu^2}\frac{d\mu^{\prime 2}}{\mu^{\prime 2}} \frac{\alpha_s(\mu')}{2\pi}\gamma^{(3)}(\mu')\right]_{jk} \\&\times\text{P}\exp\left[-\int_{\mu_0^2}^{\mu^2}\frac{d\mu^{\prime 2}}{\mu^{\prime 2}} \frac{\alpha_s(\mu')}{2\pi}\gamma^{(5)}(\mu')\right]_{ki} \ , \label{eq:matchc}
\end{split}
\end{equation}
where $\mu_0$ is a parameter that represents the scale where the behavior for $q_T\to 0$~\cite{Chang:2025kgq} takes over.
In the range $50$~GeV to $1$~TeV we observe a decrease by about $20\%$ for ${\cal C}_{qq}(\mu)$ from this formula. This effect needs to be taken into account for performing precision calculations. 

\begin{figure}[htbp]
  \begin{center}
   \includegraphics[width=0.48\textwidth]{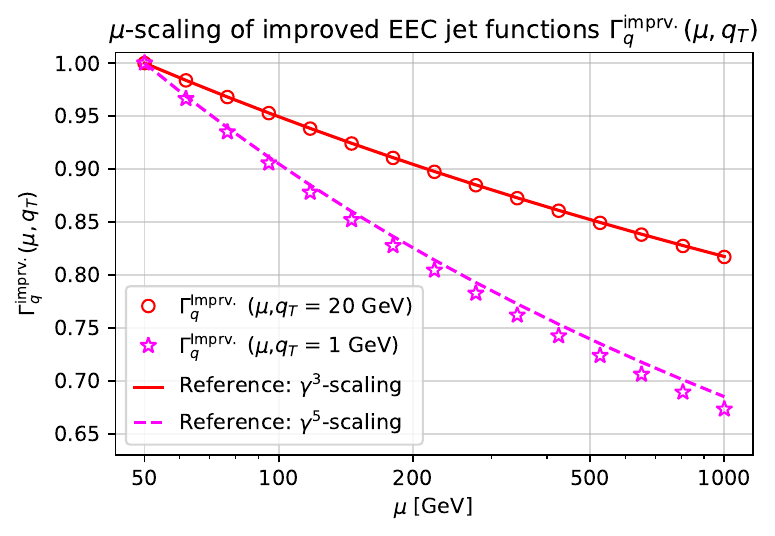}
   \vspace*{2mm}
\caption{Scaling behaviors for the improved unintegrated EEC jet function $\Gamma_q(\mu,q_T)$ in the non-perturbative and perturbative regions with different transverse momenta, $q_T=1,20$~GeV. To illustrate the scaling behavior, both curves are normalized to those at $\mu=50~\rm GeV$. }
  \label{fig:cmu}
 \end{center}
  \vspace{-5.ex}
\end{figure}
In addition, we also need to adopt a matching scheme, linking the low-$q_T$ ($<\Lambda$) region to the high-$q_T$ ($>\Lambda$) region, where $\Lambda$ is the parameter defining the nonperturbative part of the EEC jet function introduced above. A simple matching can be achieved as follows,
\begin{equation}
    \Gamma_i^{\textrm{match}}(\mu,q_T)=\Gamma_j^{\rm LLA}(\mu,q_T)\left(1+\left({\cal C}_{ji}(\mu)-1\right)e^{-{q_T^2\over  \Lambda_i^2}}\right) \ ,
\end{equation}
with the same parameter $\Lambda_i$ as used in the nonperturbative part in Eq.~(\ref{eq:nonpert}). In Fig.~\ref{fig:cmu}, we show the comparison of the EEC jet functions at $q_T=1\rm GeV$ and $q_T=20\rm GeV$ as functions of $\mu$. 
For the numerical calculations, we used the moments of globally fitted single-hadron fragmentation functions as the input of $\Gamma'(\mu_{\rm init})$ at $\mu_{\rm{init}}=2$ GeV to calculate $\Gamma'_q(\mu_{\rm init})=0.246$ and $\Gamma'_g(\mu_{\rm init})=0.176$~\cite{Gao:2024nkz}, so that $\Gamma_i(\mu_{\rm init})=1-\Gamma'_i(\mu_{\rm init})=(0.754,0.824)$. We have also chosen $\Lambda_g\approx \Lambda_q=4.37$~GeV, which gives a reasonable description of the near-side EECs in $e^+e^-$ annihilation, see the next section. The different scaling behavior is clearly visible. As a reference, we also plot the scaling behaviors governed by $\gamma^{(3)}$ and $\gamma^{(5)}$, respectively. In the following section, we will apply the above improved EEC jet functions to describe the EECs in $e^+e^-$ annihilation.

\section{EECs in $e^+e^-$ Annihilation}

EECs in $e^+e^-$ annihilation were first studied in the classic papers by Basham {\it et al.}~\cite{Basham:1977iq,Basham:1978bw,Basham:1978zq}. In order to study the angular integrated EECs and small-angle EECs, we follow the formalism developed for hadron production in $e^+e^-$ annihilation, for which factorization in terms of fragmentation functions has been shown~\cite{Altarelli:1979kv}. Our goal is to keep the IR behavior for these observables and demonstrate factorization in terms of the EEC jet functions introduced in the previous section.

\subsection{Integrated EECs}

Let us start with the differential cross sections for inclusive-hadron and hadron pair productions in $e^+e^-$ annihilation, which have been computed in perturbation theory 
previously, see, e.g., Refs.~\cite{Altarelli:1979kv,deFlorian:2003cg}. For single-hadron production, we have 
\begin{equation}
    \frac{d\sigma(e^+e^-\to h+X)}{dz}=\sigma_0\int_z \frac{dx}{x}D_i^h(z/x,\mu)H_i(x,\mu) \ ,\label{eq:eeh}
\end{equation}
where $z$ is the fraction of the virtual photon's momentum $q$ carried by the hadron and as before $D_i^h(z,\mu)$ represents the fragmentation function for parton $i$ to hadron $h$. In the above equation, $\sigma_0$ is the leading-order cross section and $H_i$ is the partonic one. Up to one-loop order, we write
\begin{eqnarray}
    H_q&=&\delta(1-x)+\frac{\alpha_s}{2\pi} C^{e^+e^-}_q(x) \ ,\\
    H_g&=&\frac{\alpha_s}{2\pi}C^{e^+e^-}_g(x) \ ,
\end{eqnarray}
where $C^{e^+e^-}_q$ and $C^{e^+e^-}_g$ are listed in the Appendix for convenience. 

From the above, we can derive the one-point energy correlator,
\begin{equation}
    \Sigma_1=\sum_i\frac{E_i}{Q}=\frac{1}{\sigma_{\textrm{tot}}}\int dz \frac{1}{2}z \sum_h\frac{d\sigma^h}{dz}  \ ,
\end{equation}
where $d\sigma^h$ is a short-hand notation for $d\sigma(e^+e^-\to h+X)$. 
By applying energy conservation of the fragmentation functions, $\int d z \sum_hD_i^h(z,\mu)\equiv 1$, it has long been realized that the above $\Sigma_1\equiv 1$ as well. This can be easily checked with the following results,
\begin{eqnarray}
&&    \sigma_{\textrm{tot}}=\sigma_0\left(1+\frac{\alpha_s}{\pi}\right)\ ,\\
&&    \int dx C^{e^+e^-}_q(x)=2\ ,\\
&&    \int dx x \left(C^{e^+e^-}_q(x)+C^{e^+e^-}_g(x)\right)=2 \ .
\end{eqnarray}

We now turn to two-particle productions, where the differential cross section contains two components,
\begin{eqnarray}
    \frac{d\sigma^{h_1h_2}}{dz_1dz_2}=\left.\frac{d\sigma^{h_1h_2}}{dz_1dz_2}\right|_{\mathrm{hom.}}+\left.\frac{d\sigma^{h_1h_2}}{dz_1dz_2}\right|_{\mathrm{inhom.}} \ ,
\end{eqnarray}
associated with homogeneous di-hadron fragmentation and two separate individual fragmentations, respectively. 
The homogeneous term takes the same form as above for single-particle production,
\begin{equation}
\left.\frac{d\sigma^{h_1h_2}}{dz_1dz_2}\right|_{\mathrm{hom.}}=\sigma_0\int_{z_1+z_2} \frac{dx}{x^2}I\!\!D_i^{h_1h_2}\left(\frac{z_1}{x},\frac{z_2}{x},\mu\right)H_i(x,\mu) \ ,    
\end{equation}
where $I\!\!D_i^{h_1h_2}(\frac{z_1}{x},\frac{z_2}{x},\mu)$ is again the di-hadron fragmentation function, and $H_i$ has been defined above.

The integrated two-point EEC is calculated from the above differential cross section as
\begin{equation}
    \Sigma_2\equiv\sum_{i\neq j}\frac{E_iE_j}{Q^2}=\frac{1}{\sigma_{\textrm{tot}}}\int dz_1dz_2 \frac{1}{2}z_1 z_2\sum_{h_1,h_2}\frac{d\sigma^{h_1h_2}}{dz_1dz_2}  \ .
\end{equation}
At the leading order, we have
\begin{equation}
    \Sigma_2^{(0)}=\frac{1}{2}\left(1+\Gamma_q(Q)\right) \ ,
\end{equation}
where the first term comes from the inhomogeneous contribution and the second from the homogeneous one. 

From the coefficients listed in the Appendix, we can also compute the one-loop corrections. First, the homogeneous term is easy to evaluate, we have 
\begin{eqnarray}
    \Sigma_2^{\textrm{hom.}(1)}&=&\frac{\sigma_0}{2\sigma_{\textrm{tot}}}\frac{\alpha_s}{2\pi}\left[-\left(-\frac{1}{\hat \epsilon}+\ln\frac{Q^2}{\mu^2}\right)\gamma_{qq}^{(3)}\Gamma_q(\mu)\right.\nonumber\\
    &&-\left(-\frac{1}{\hat \epsilon}+\ln\frac{Q^2}{\mu^2}\right)\gamma_{gq}^{(3)}\Gamma_g(\mu)\nonumber\\
    &&\left. +\frac{131}{12}\Gamma_q(\mu)-\frac{71}{36}\Gamma_g(\mu)\right] \ ,
\end{eqnarray}
where $1/\hat \epsilon\equiv 1/\epsilon +\log(4\pi)-\gamma_E$. The first two terms contain collinear divergences that are related to the renormalization of $\Gamma_q(\mu)$ at one-loop order. The last line represents the NLO corrections, and the coefficients are obtained from $\int dx x^2 C^{e^+e^-}_{q,g}(x)$, respectively.

Similarly, we also compute the inhomogeneous term, which has a collinear divergence as well:
\begin{equation}
    \Sigma_2^{\textrm{inhom.}(1)}=\frac{\sigma_0}{2\sigma_{\textrm{tot}}}\frac{\alpha_s}{2\pi}C_F\left[-\frac{1}{\hat \epsilon}\frac{3}{2}+\frac{3}{2}\ln\frac{Q^2}{\mu^2}-\frac{89}{24}\right] \ .\label{eq:eeinhomo}
\end{equation}
The above is derived from the two-parton differential cross sections; for details, see the Appendix. In particular, in dimensional regularization, the collinear divergence comes from the following angular distribution at small angle:
\begin{equation}
    \left.\frac{d\Sigma_2}{d\zeta}\right|_{\zeta\to 0}=\frac{\sigma_0}{2\sigma_{\textrm{tot}}}\frac{\alpha_s}{2\pi}C_FN_\epsilon\left[\frac{3}{2}\frac{1}{\zeta^{1+\epsilon}}+{\rm finite~term}\right] \ ,
\end{equation}
where  $\zeta=(1-\cos\theta)/2$ with $\theta$ the angle between the two hadrons. Furthermore, $N_\epsilon=\frac{\Gamma(1+\epsilon)\Gamma^2(1-\epsilon)}{\Gamma(1-2\epsilon)}(\frac{Q^2}{4\pi\mu^2})^{-\epsilon}$. The ``finite term'' part contains the regular contribution when $\zeta\to 0$, including both real and virtual pieces. Clearly, the $1/\zeta$ behavior corresponds to a collinear divergence that needs to be taken care of by proper renormalization of the inhomogeneous part of $\Gamma_q(Q)$ at this order. The final result for $\Sigma_2$ can be written as
\begin{eqnarray}
  \frac{\sigma_{\textrm{tot}}}{\sigma_0} \Sigma_2&=&\frac{1}{2}\left(1+\frac{\alpha_s}{2\pi}\left(-\frac{89}{24}\right)C_F\right)\\
    &+&\frac{1}{2}\left(\Gamma_q(Q)+\frac{\alpha_s}{2\pi}\left[\frac{131}{12}\Gamma_q(Q)-\frac{71}{36}\Gamma_g(Q)\right] \right) \ . \nonumber
\end{eqnarray}
At one-loop order, $\sigma_{\textrm{tot}}=\sigma_0(1+\alpha_s/\pi)$ (we note that $\sigma_0/\sigma_{\textrm{tot}}$ has been calculated perturbatively up to NNLO~\cite{Kardos:2018kqj}). It is interesting to note that asymptotically, $Q^2\to \infty$, we have $\Gamma_q(Q)\sim\Gamma_g(Q)\to 1$ and $\Sigma_2\to 1$. This provides an important cross check of our above results, in particular, for the inhomogeneous contributions.

\begin{figure}[htbp]
  \begin{center}
   \includegraphics[scale=0.6]{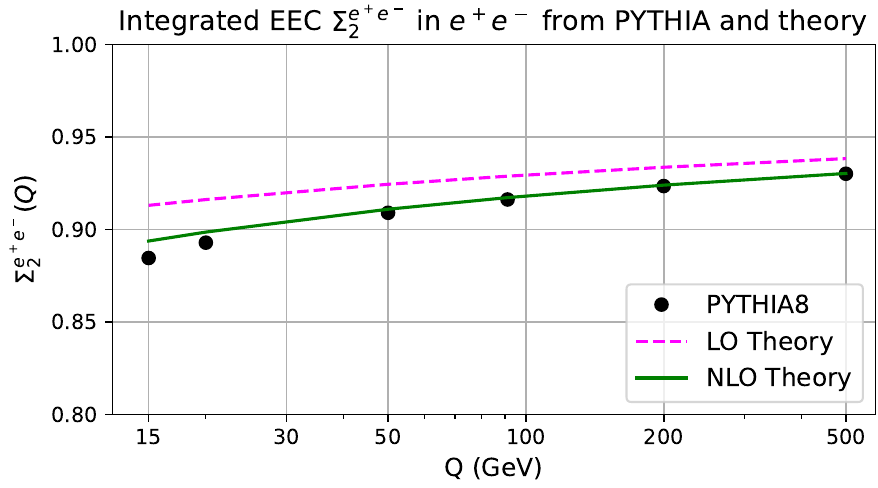} 
\caption{Integrated EEC in $e^+e^-$ annihilation: theory predictions compared to PYTHIA8 simulations. }
  \label{fig:integratedEEC}
 \end{center}
  \vspace{-5.ex}
\end{figure}

In Fig.~\ref{fig:integratedEEC}, we show the comparison between the above LO and NLO results to PYTHIA8 simulations for the integrated EEC in $e^+e^-$ annihilation.
We observe striking overall agreement between NLO theory and PYTHIA8, which becomes excellent as soon as $Q>20$~GeV.

\subsection{Near-Side EEC \label{sec:nearEEC}}

Focusing on the near-side EEC, we have, at the leading order,
\begin{equation}
    \left.2\frac{d\Sigma_2^{(0)}}{\pi Q^2d\zeta}\right|_{\zeta\to 0}=
    \Gamma_q^{(0)}(\mu,q_T) \ ,\label{eq:NearEEC}
\end{equation}
where $q_T^2=\zeta Q^2$, again with $\zeta=(1-\cos(\theta))/2\approx \theta^2/4$ at small angle. The quark carries energy $E_q=\overline Q=Q/2$, so that $q_T=2 E_q\sqrt\zeta$. Equation~(\ref{eq:NearEEC}) defines the unintegrated EEC jet function $\Gamma_q^{(0)}(\mu,q_T)$, describing the distribution in opening angle $\theta\propto q_T$ for di-hadron fragmentation. At small angle the unintegrated EEC jet function is nonperturbative. Its normalization is given by
\begin{equation}
\Gamma_q^{(0)}(\mu)=\int d^2q_T \Gamma_q^{(0)}(\mu,q_T) \ ,    
\end{equation}
where the left side is the integrated EEC jet function. Following the same strategy as for single-inclusive hadron production in $e^+e^-$ annihilation~\cite{Altarelli:1979kv}, one can compute one-loop corrections for the near-side EEC from the EEC jet function. A similar factorization like Eq.~(\ref{eq:eeh}) will follow,
\begin{equation}
    \left.2\frac{d\Sigma_2^{e^+e^-}}{\pi Q^2d\zeta}\right|_{\zeta\to 0}={\sigma_0\over\sigma_{\textrm{tot}}}\int {dx}{x^4}
    H_i\left(x,{Q\over\mu}\right)\Gamma_i\left(\mu,q_T\right) \ ,\label{eq:S2fact}
\end{equation}
where $E_i=x\overline Q$ and $q_T^2=\zeta x^2Q^2$. Again, we make sure the EEC is properly normalized. In terms of the Fourier transform to $b_T$ space we have
\begin{eqnarray}
     \left.2\frac{d\Sigma_2^{e^+e^-}}{\pi Q^2d\zeta}\right|_{\zeta\to 0}&=&{\sigma_0\over\sigma_{\textrm{tot}}}\int {dx} {x^2}H_i\left(x,{Q\over\mu}\right)\nonumber\\
    &\times &\int \frac{b_Td\bar b_T}{(2\pi)}J_0({\sqrt{\zeta}Q\bar b_T})\widetilde{\Gamma}_i\left(\mu,{\bar b_T\over x}\right) \ ,\;\;\;\;\;\;\label{eq:factorizationbt}
\end{eqnarray}
where the EEC jet function in $b_T$ space is defined as
\begin{equation}
    \widetilde{\Gamma}_i(\mu,b_T)=\pi\int d q_T^2 J_0(q_Tb_T)\Gamma_i(\mu,q_T)\ .
\end{equation}
Using this definition and rearranging Eq.~(\ref{eq:S2fact}), we obtain the $b_T$-space factorization formula 
\begin{equation}
    \left.2\frac{d\Sigma_2^{e^+e^-}}{\pi Q^2d\zeta}\right|_{\zeta\to 0}={\sigma_0\over\sigma_{\textrm{tot}}}
    \int \frac{b_Tdb_T}{(2\pi)}J_0({\sqrt{\zeta}Qb_T})\widetilde{\Sigma}_2(Q,b_T) \ ,
\end{equation}
where
\begin{equation}
    \widetilde{\Sigma}_2(Q,b_T)\equiv\int {dx} {x^2}H_i\left(x,{Q\over \mu}\right)\widetilde{\Gamma}_i\left(\mu,{b_T\over x}\right) \ .\label{eq:factorizationbtee}
\end{equation}
The leading order is simple since $H_q^{(0)}(x)=\delta(1-x)$, and we have
\begin{equation}
    \widetilde{\Sigma}_2^{(0)}(Q,b_T)=\widetilde{\Gamma}^{(0)}_q(Q,b_T) \ .
\end{equation}
At one-loop order, we need to take into account real and virtual contributions. As before, we can again classify these contributions into homogeneous and inhomogeneous ones. For the homogeneous contribution, we can make use of the known results for single-parton differential cross sections. For example, for the quark EEC contribution we integrate out the phase space of the radiated gluon and antiquark. The differential cross section for the quark and gluon distributions in $e^+e^-\to q\bar q g$ can be written as
\begin{eqnarray}
    \frac{d\sigma^q}{dx}&=&\sigma_0\frac{\alpha_s}{2\pi}\left[\left(-\frac{1}{\hat \epsilon}+\ln\frac{Q^2}{\mu^2}\right)\hat P_{qq}(x)+C^{e^+e^-}_q(x)\right],\;\;\;\;\;\;\\
    \frac{d\sigma^g}{dx}&=&\sigma_0\frac{\alpha_s}{2\pi}\left[\left(-\frac{1}{\hat \epsilon}+\ln\frac{Q^2}{\mu^2}\right)\hat P_{gq}(x)+C^{e^+e^-}_g(x)\right].\;\; \;\;   \;\;
\end{eqnarray}
From this, we find for the EEC at small angle,
\begin{eqnarray}
&&   \left.2\frac{d\Sigma_2^{(1)}}{\pi Q^2d\zeta}\right|_{\zeta\to 0}={\sigma_0\over\sigma_{\textrm{tot}}}
   \frac{\alpha_s}{2\pi}\int dx  x^4 \\%\Gamma_q^{(0)}(E_q,q_T)\\
  &&\times  \left\{  \Gamma_q^{(0)}(\mu,q_T)\left[\left(-\frac{1}{\hat \epsilon}+\ln\frac{Q^2}{\mu^2}\right)\hat P_{qq}(x)+C^{e^+e^-}_q(x)\right] \right.\nonumber\\
  &&\left. +  \Gamma_g^{(0)}(\mu,q_T)\left[\left(-\frac{1}{\hat \epsilon}+\ln\frac{Q^2}{\mu^2}\right)\hat P_{gq}(x)+C^{e^+e^-}_g(x)\right] \right\}  \ . \nonumber
\end{eqnarray}
Fourier transforming the above to $b_T$-space, we obtain the homogeneous contribution to $\widetilde{\Sigma}_2$ at one-loop order,
\begin{eqnarray}
&&    \widetilde{\Sigma}_2^{\textrm{hom.}(1)}=\frac{\alpha_s}{2\pi}\int dx x^2\\
&&\times \left\{\widetilde{\Gamma}_q^{(0)}(\mu,{b_T\over x})\left[\left(-\frac{1}{\hat \epsilon}+\ln\frac{Q^2}{\mu^2}\right)\hat P_{qq}(x)+C^{e^+e^-}_q(x)\right]\right.\nonumber\\
    &&\left. +\widetilde{\Gamma}_g^{(0)}(\mu,{b_T\over x})\left[\left(-\frac{1}{\hat \epsilon}+\ln\frac{Q^2}{\mu^2}\right)\hat P_{gq}(x)+C^{e^+e^-}_g(x)\right]\right\} \ . \nonumber
\end{eqnarray}
For the inhomogeneous contribution two partons are close to each other and each of them separately fragments into a hadron; see previous subsection.
We have
\begin{equation}
\left.2\frac{d\Sigma_2^{(1)}}{\pi Q^2d\zeta}\right|_{\zeta\to 0}={\sigma_0\over\sigma_{\textrm{tot}}}
\frac{\alpha_s}{2\pi}\left(\gamma_{qq}^{(3)}+\gamma_{gq}^{(3)}\right)\frac{1}{\zeta^{1+\epsilon}}\ .
\end{equation}
It is again convenient to express the final result in Fourier transform $b_T$-space,
\begin{equation}
\widetilde{\Sigma}_2^{\textrm{inhom.}(1)}=\frac{\alpha_s}{2\pi}\left(\gamma_{qq}^{(3)}+\gamma_{gq}^{(3)}\right)\left(-\frac{1}{\hat \epsilon}+\ln\frac{\mu_b^2}{\mu^2}\right)\ .
\end{equation}
Adding the homogeneous and inhomogeneous contributions, we have
\begin{eqnarray}
    \widetilde{\Sigma}_{2,\textrm{tot}}^{(1)}&=&\frac{\alpha_s}{2\pi}\left\{\int dx x^2\left(-\frac{1}{\hat \epsilon}+\ln\frac{Q^2}{\mu^2}\right)\left[\widetilde{\Gamma}_q^{(0)}\left(\mu,{b_T\over x}\right)\right.\right.\nonumber\\
    &&\times \hat P_{qq}(x)\left.+\widetilde{\Gamma}_g^{(0)}\left(\mu,{b_T\over x}\right)
    \hat P_{gq}(x)\right]\nonumber\\
&& +\left.  
\left(\gamma_{qq}^{(3)}+\gamma_{gq}^{(3)}\right)\left(-\frac{1}{\hat \epsilon}+\ln\frac{\mu_b^2}{\mu^2}\right) \right\}\nonumber\\
&&+\frac{\alpha_s}{2\pi}\int dx x^2\left[\widetilde{\Gamma}_q^{(0)}\left(\mu,{b_T\over x}\right)C_q^{e^+e^-}(x)\right.\nonumber\\
&&\left.+\widetilde{\Gamma}_g^{(0)}\left(\mu,{b_T\over x}\right)C_g^{e^+e^-}(x)\right]\ ,\label{eq:oneloopbtee}
\end{eqnarray}
where the first term is just part of the one-loop result for $\widetilde{\Gamma}_q^{(1)}(Q,b_T)$ in Eq.~(\ref{eq:oneloopbt0}) of Sec.~\ref{sec:unintEEC}.
The above result demonstrates the following factorization formula at one-loop order:
\begin{equation}
    \widetilde{\Sigma}_2^{(1)}(Q,b_T)=\int dx x^2 \left[H_q^{(0)}\widetilde{\Gamma}_q^{(1)}+H_i^{(1)}(x)\widetilde{\Gamma}_i^{(0)}\right] \ ,
\end{equation}
with the hard coefficients
\begin{eqnarray}
    H_q&=&\delta(1-x)+\frac{\alpha_s}{2\pi} C^{e^+e^-}_q(x) \ ,\\
    H_g&=&\frac{\alpha_s}{2\pi}C^{e^+e^-}_g(x) \ .
\end{eqnarray} 
To simplify the above expressions, we have set the factorization scale to $\mu=Q$. The complete scale dependence can be restored from the above one-loop results (\ref{eq:oneloopbtee}) for the factorization formula (\ref{eq:factorizationbtee}). It is important to note that the scale dependence is the same as that for single particle production in $e^+e^-$ annihilation. 

To carry out resummation of logarithms in $b_T$, we can apply the resummed 
EEC jet functions $\Gamma_i$ derived in Sec.~\ref{sec:match}, for instance within LLA in Eq.~(\ref{eq:nonpert}):
\begin{equation}
    \widetilde{\Sigma}^{\textrm{res,LLA}}_2(Q,b_T)=\int dx x^2 {\bf H}\left(x,{Q\over \mu}\right)\cdot {\bf \Gamma}_{\textrm{res}}^{\textrm{LLA}}\left(\mu,{b_T\over x}\right) \ .\label{eq:resum_ee1}
\end{equation}
As shown in Ref.~\cite{Guo:2025zwb}, in LLA, the above result will lead to the same asymptotic expansion as given in Ref.~\cite{Dixon:2019uzg}. With hard factors computed at one-loop order, we should be able to expand the resummation result beyond the LLA, where additional contributions have to be taken into account including full evolutions at NNLL, following the
lines in Ref.~\cite{Dixon:2019uzg}. We will come back to this in a future publication.

\subsection{Comparison to Experimental Data}

Applying the factorization result derived above with the improved EEC jet function from the previous section, we obtain the final result for the EEC at small angle in $e^+e^-$ annihilation, 
\begin{eqnarray}
    &&\left.2\frac{d\Sigma_2^{e^+e^-}}{\pi Q^2d\zeta}\right|_{\zeta\to 0}\nonumber\\
   && ={\sigma_0\over\sigma_{\textrm{tot}}}\int {dx}{x^4}
    H_i\left(x,{Q\over\mu}
    \right)
    \left(1+\left({\cal C}_{ji}(\mu)-1\right)e^{-{q_T^2\over  \Lambda_i^2}}\right)\nonumber\\
    &&~~~~~~~\times \int\frac{d^2b_T}{(2\pi)^2}e^{iq_T\cdot b_T}e^{-\Lambda_jb_T}\sum_k\Gamma_k(\mu_{b*})\nonumber\\
    &&~~~~~~~~~\times\text{P}\exp
    {\left[-\int_{\mu_{b*}^2}^{\mu^2}\frac{d\mu^{\prime 2}}{\mu^{\prime 2}} \gamma^{(3)}(\mu')\right]_{kj}} \ , 
\label{eq:eefinalimproved}    
\end{eqnarray}
where $q_T^2=\zeta Q^2$, $\mu_{b*}=2e^{-\gamma_E}/b_*$ with $b_*=b_T/\sqrt{1+b_T^2/b_{\textrm{max}}^2}$ and  where we set $b_{\textrm{max}}=1.5~\rm GeV^{-1}$. In the above equation, the factor ${\cal C}_{ji}(\mu)$ represents the matching between the perturbative and nonperturbative regions which have different scaling behaviors, see, Sec.~\ref{sec:match} and Eq.~(\ref{eq:matchc}). We have also set ${\cal C}_{ji}(\mu=20\,{\rm GeV})=1$ at an adopted initial scale for the matching. The second and third lines of Eq.~(\ref{eq:eefinalimproved}) represents the LLA resummation for the EEC jet function. In total, we only have two non-perturbative parameters $\Lambda_{q.g}$. Because the EEC in $e^+e^-$ annihilation is more sensitive to the quark EEC jet function, we choose $\Lambda_{g}\approx \Lambda_q$ for simplicity and 
also use the same value for all quark flavors.

Before we compare the above equation to PYTHIA8 simulations, %
we observe that the NLO correction is quite sizable, especially at low $q_T$, where it reaches more than $20\%$. This can be understood by computing the moments of the Wilson coefficients $C_{q,g}^{e^+e^-}$ and taking the $q_T\to 0$ limit of Eq.~(\ref{eq:eefinalimproved}),
\begin{eqnarray}
    \left.K^{\rm NLO}\right|_{q_T\to 0}&\approx& 1+\frac{\alpha_s}{2\pi}\left(\int dx x^4\left[C_q^{e^+e^-}+C_g^{e^+e^-}\right] -2\right)\nonumber\\
    &=& 1+\frac{\alpha_s}{2\pi}
    \frac{5491}{450}
    \ ,
\end{eqnarray}
where we have made the approximation $\Gamma_q\approx \Gamma_g\approx 1$. The $K$-factor decreases with increasing $q_T$ because the NLO corrections push the distribution to smaller $q_T$ due to the integral of the hard coefficients $C_{q/g}^{e^+e^-}(x)$ multiplied by the EEC jet functions.

\begin{figure}[htbp]
  \begin{center}
   \includegraphics[width=0.484\textwidth]{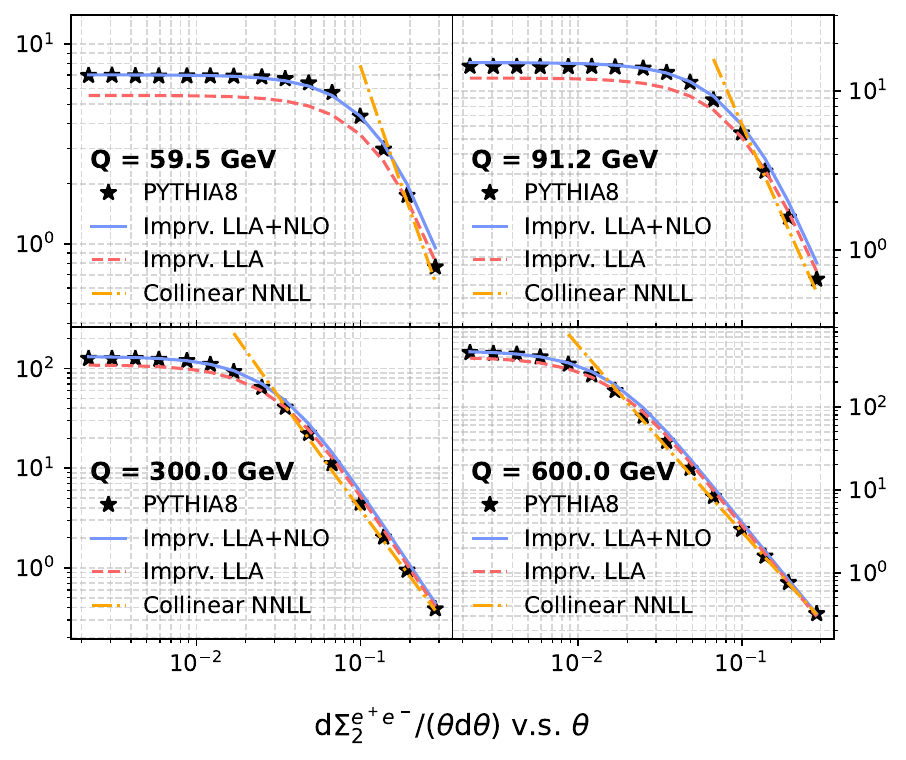} 
\caption{Comparison of $\text{d}\Sigma_2^{e^+e^-}/(\theta\text{d}\theta)$ in Eq.~(\ref{eq:eefinalimproved}) to PYTHIA8 simulations. We have imposed a normalization factor $N_{\rm{PY}}=0.85$ to our results, which reflects the fact that the PYTHIA8 simulations are about 15\%  lower than the experimental data.}
  \label{fig:fiteecsim}
 \end{center}
  \vspace{-5.ex}
\end{figure}

\begin{figure}[htbp]
  \begin{center}
   \includegraphics[width=0.49\textwidth]{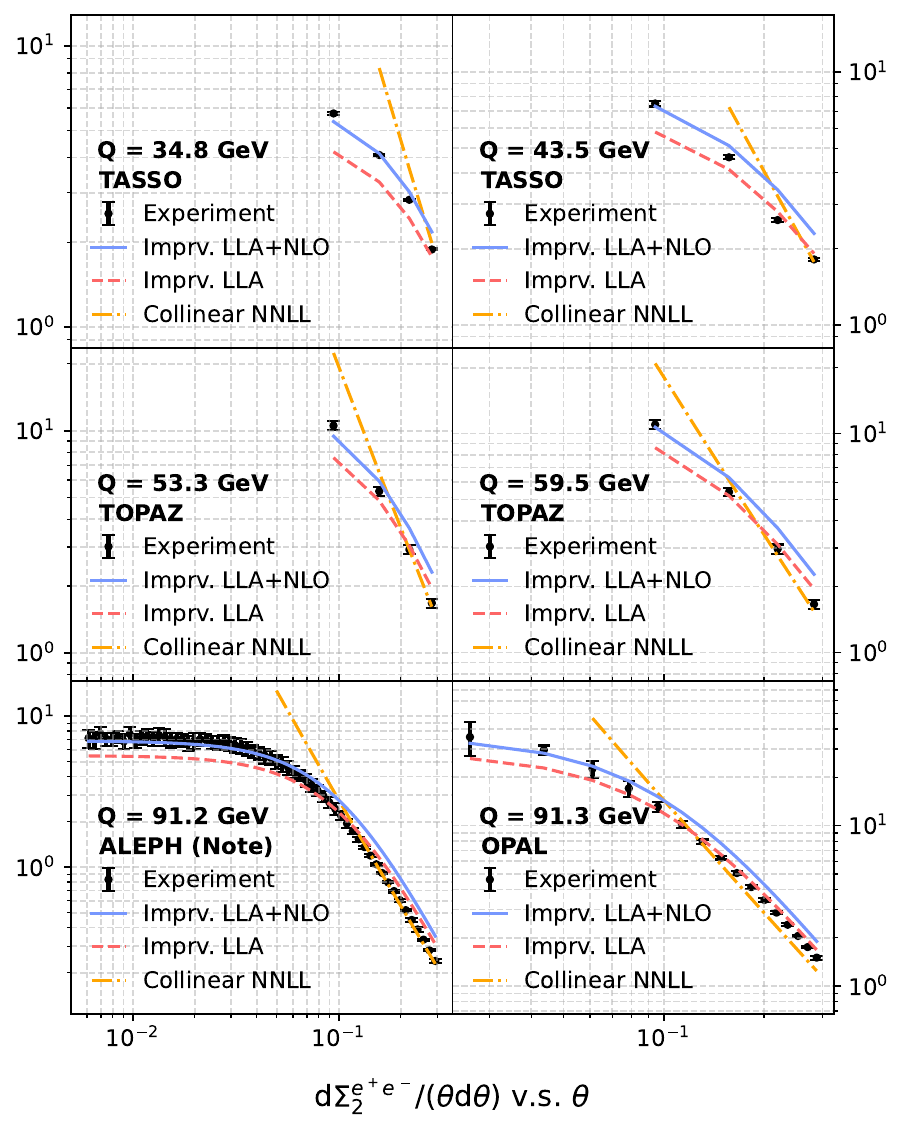} 
\caption{Comparison of $\text{d}\Sigma_2^{e^+e^-}/(\theta\text{d}\theta)$ in Eq.~(\ref{eq:eefinalimproved}) to the experimental data for $Q>30$ GeV. We include a conversion factor of $4/9$ in the comparison to the latest ALEPH data (bottom left), which are for all charged hadrons, and another factor $1/2$ because ALEPH counts each particle pair $(i,j)$ only once rather than twice. Including these two factors, we noted a $15\%$ excess of the new ALEPH data relative to the OPAL data at the same energy. This could be due to a correction to our ``naive'' $4/9$ factor. A further normalization factor of $1/1.15$ is therefore applied when comparing to ALEPH. We note that no extra normalization factors are needed for any of the other data sets.}
  \label{fig:fiteecexp}
 \end{center}
  \vspace{-5.ex}
\end{figure}

We now turn to the non-perturbative parameters $\Lambda_{q,g}$ in Eq.~(\ref{eq:eefinalimproved}). In general, these do not affect the distributions at moderate and large $q_T\gg \Lambda_{\rm QCD}$, whereas they play an important role at low $q_T$. Our numerical results show that the low-$q_T$ plateau of the EEC approximately scales as $1/\Lambda_q^{1.6}$, so that the plateau may be used to determine the actual values of $\Lambda_{q,g}$. In Fig.~\ref{fig:fiteecsim}, we compare our improved LLA result of Eq.~(\ref{eq:eefinalimproved}) to PYTHIA8 simulations for different energies. We find that the PYTHIA8 simulations are approximately $15\%$ lower than the experimental measurements (see also Fig.~\ref{fig:fiteecexp}). To account for this discrepancy, we introduce an overall normalization factor $N_{\rm PY}=0.85$ to align our results with the PYTHIA8 simulations, as illustrated in the figure. With this normalization applied, we tune the nonperturbative parameters to reproduce the simulations and obtain $\Lambda_g \simeq \Lambda_q = 4.37~\mathrm{GeV}$.

Figure~\ref{fig:fiteecexp} shows the comparison between our results based on Eq.~(\ref{eq:eefinalimproved}) and experimental data for a wide range of energies~\cite{TASSO:1987mcs,AMY:1988yrv,TOPAZ:1989yod,ALEPH:1990vew,L3:1991qlf,L3:1992btq,DELPHI:1990sof,OPAL:1990reb,OPAL:1991uui}. We have also included a comparison to the latest ALEPH data for charged hadrons~\cite{Electron-PositronAlliance:2025fhk}. If we assume roughly equal numbers of produced positive, negative and neutral particles, the charged-hadron {\it single-particle} cross section would be expected to be about $2/3$ of that for all hadrons. For EEC observables, by simple counting, this would result in a factor of $(2/3)^2=4/9$~\footnote{For the charge track measurements of~\cite{Electron-PositronAlliance:2025fhk}, a different factorization should apply to properly account for this difference, see, Ref.~\cite{Jaarsma:2025tck}.}. The comparison to the recent ALEPH data indeed indicates this difference. We have also verified this factor by a PYTHIA simulation. We would like to emphasize that, although the pure LLA results can also give a qualitative description of the EECs at different energies, with the improvement by the factor ${\cal C}_{ij}$ the agreement between theory and simulations becomes much better. This highlights the importance of including the scaling behavior constraints for the EEC jet functions at low $q_T$~\cite{Chang:2025kgq}.  

The comparisons in the above two figures show that Eq.~(\ref{eq:eefinalimproved}) can give a reasonable description of the EECs in $e^+e^-$ annihilation with just one parameter. Of course, theory improvements are still needed, including, for example, the NNLO corrections and/or full evolution of the EEC jet functions. While the former can be achieved by following the derivations of higher order calculations of single-inclusive hadron production~\cite{Rijken:1996ns,Mitov:2006wy,He:2025hin}, the latter may take extra efforts in theory developments. We plan to come back to these issues in a future publication.

\section{EECs in Deep Inelastic Scattering}

Lepton-nucleon scattering offers further exciting opportunities for studying EECs. We devote this section to EEC observables in DIS processes. Because EECs are constructed from hadrons in the final state, we can base our analysis on the well established semi-inclusive hadron production in DIS (SIDIS), $\ell (p_\ell)+N(p_N)\to \ell'(p_\ell')+h(p_h)+X$, where we have indicated the momenta. The SIDIS differential cross section can be written as (see, e.g., Refs.~\cite{Altarelli:1979kv,Furmanski:1981cw,deFlorian:1997zj})
\begin{eqnarray}
    \frac{d\sigma^h}{dx_Bdydz_h}&=&\frac{2\pi\alpha^2}{Q^2}\left[\frac{1+(1-y)^2}{y}2F_1^h(x_B,z_h,Q^2)\nonumber\right.\\
    &&~~~\left.+\frac{2(1-y)}{y}F_L^h(x_B,z_h,Q^2)\right] \ ,
\end{eqnarray}
where $x_B=Q^2/2P_N\cdot q$ and $y=(E_\ell-E_\ell')/E_\ell$ are the usual DIS kinematic variables, and $z_h=p_h\cdot P_N/q\cdot P_N$ represents the momentum fraction of the final-state hadron with respect to the incoming virtual photon. The SIDIS structure functions $F_{I=T,L}^h$ can be factorized into the parton distribution functions $f_i$ of the nucleon and fragmentation functions $D_j^h$  for the final-state hadron,
\begin{eqnarray}
    F_I(x_B,z_h)&=&\sum_{i,j}\int \frac{d\xi}{\xi}\frac{d\hat\xi}{\hat\xi}f_i\left(\frac{x_B}{\xi},\mu\right)D_j^h\left(\frac{z_h}{\hat\xi},\mu\right)\nonumber \\
    &&\times C_{ji}^I(\xi,\hat\xi,\mu) \ ,
\end{eqnarray}
where the sums run over parton species. 
The hard coefficients $C_{ji}^I$ can be computed in perturbative QCD. We list the corresponding first-order expressions in the Appendix for reference. 

From the above, we can define the one-point energy correlator for SIDIS:
\begin{equation}
    \Sigma_1^{\rm DIS}=\frac{1}{\sigma_{\textrm{tot}}}\sum_h\int d\sigma \frac{p_h\cdot p_N}{q\cdot p_N}\ ,
\end{equation}
where we sum over hadrons in the final state. Starting from the SIDIS differential cross section, we write 
\begin{equation}
    \Sigma_1^{\rm DIS}=\frac{1}{d\sigma(x_B,y)}\sum_h\int dz_h z_h\frac{d\sigma^h}{dx_Bdydz_h} \ .
\end{equation}
Because of momentum conservation, we immediately conclude that $\Sigma_1^{\rm DIS}\equiv 1$~\cite{Altarelli:1979kv}. One can also study the angular distribution of the above one-point energy correlator with respect to the incoming lepton, which in DIS experiments is referred to as the angular distribution of energy flow~\cite{H1:1995tux}. A TMD factorization should be applied to describe the relevant physics~\cite{Nadolsky:1999kb,Li:2020bub}.  

\subsection{Integrated EEC in SIDIS}

For the EEC observable in SIDIS, we define
\begin{equation}
    \Sigma_2^{\rm DIS}=\frac{1}{\sigma_{\textrm{tot}}}\sum_{h_1h_2}\int d\sigma \frac{p_{h_1}\cdot p_Np_{h_2}\cdot p_N}{(q\cdot p_N)^2}\ , \label{eq:eecdisdef0}
\end{equation}
where the weight factors represent the fractions of the energy of the incoming photon carried by the two final state hadrons. They are frame-independent kinematic variables and can hence be determined in any frame. The physics arguably becomes most transparent in the Breit frame, as will be discussed in the next subsection for the angular distributions. 

The EEC $\Sigma_2^{\rm DIS}$ can be computed from the differential cross section for di-hadron production in DIS,
\begin{equation}
    \Sigma_2^{\rm DIS}=\frac{1}{\sigma_{\textrm{incl.}}}\int dz_{h_1}dz_{h_2} (z_{h_1}z_{h_2})\frac{d\sigma^{h_1h_2}}{dx_Bdydz_{h_1}dz_{h_2}}\ , \label{eq:diseec0}
\end{equation}
where $z_{h_i}=p_{h_i}\cdot p_N/q\cdot p_N$ and $\sigma_{\textrm{incl.}}$ represents the inclusive differential cross section obtained by integrating out all final-state hadrons. It depends on $x_B$ and $y$. Following the calculations in the previous section, we can write down a factorization formula for the integrated EEC (omitting for brevity dependence on non-essential variables):
\begin{eqnarray}
    \Sigma_2^{{\rm{DIS}},I}&=&\frac{1}{F_I(x_B)}\int \frac{d\xi}{\xi} f_i\left(\frac{x_B}{\xi},\mu\right)\left[\Gamma_j(\mu) C_{ij}^{\textrm{DIS},I}(\xi)\right.\nonumber\\
    &&\left.~~+{\cal D}_{i}^{I}(\xi)\right] \ , \label{eq:diseec1}    
\end{eqnarray}
where, as before, $I=L,T$. The $\Gamma_j$ in the first term on the right-hand side are the EEC jet functions defined previously in Eq.~(\ref{eq:dihadron0}). As before, this term describes the homogeneous contribution, for which the two hadrons come from a single parton progenitor via a di-hadron fragmentation function. The second term represents the inhomogeneous contributions which come from individual fragmenting partons to the two hadrons. The coefficients $C_{ij}^{\textrm{DIS},I}$ and ${\cal D}_{i}^{I}$ will be derived below. We note that in the denominator of Eq.~(\ref{eq:diseec1}) we have the inclusive DIS structure functions $F_{L,T}$ which can be factorized in terms of the parton distributions:
\begin{equation}
    F_I(x_B)=\int\frac{d\xi}{\xi}C_i^I(\xi,\mu)f_i\left({x_B\over\xi}\right) \ ,
\end{equation}
where the first-order corrections to the inclusive hard coefficients are well-known in the literature,
\begin{eqnarray}
&&    C_q^{1,(1)}\nonumber\\
    &&~~~~=\delta(1-\xi)+\frac{\alpha_sC_F}{2\pi}\left[\frac{\hat P_{qq}(\xi)}{C_F}\ln\frac{Q^2}{\mu^2}+L_1(\xi)-L_2(\xi)\right.\nonumber\\
    &&~~~~\left.-\frac{3}{2}\frac{1}{(1-\xi)_+}+3+\delta(1-\xi)\left(-\frac{9}{2}-\frac{\pi^2}{3}\right)\right]\ ,
    \end{eqnarray}
    \begin{eqnarray}
    C_g^{1,(1)}&=&
    \hat P_{qg}(\xi)\left[\ln\left(\frac{Q^2}{\mu^2}\frac{1-\xi}{\xi}\right)-1\right]+\xi(1-\xi)\ ,\\
    C_q^{L,(1)}&=& 2C_F\xi \ ,\\
    C_g^{L,(1)}&=& 4\xi(1-\xi) \ ,
\end{eqnarray}
with $L_1(\xi)=(1+\xi^2)\left(\frac{\ln(1-\xi)}{1-\xi}\right)_+$ and $L_2(\xi)=\frac{1+\xi^2}{1-\xi}\ln\xi$.

The homogeneous contributions can be determined by extending the SIDIS differential cross section to a 
di-hadron differential one:
\begin{eqnarray}
    \frac{d\sigma^{h_1h_2}|_{\textrm{hom.}}}{dx_Bdydz_1dz_2}\!\!&=\!\!&\frac{2\pi\alpha^2}{Q^2}\left[\frac{1+(1-y)^2}{y}2F_1^{h_1h_2}(x_B,z_1,z_2)\nonumber\right.\\
    &&~~~\left.+\frac{2(1-y)}{y}F_L^{h_1h_2}(x_B,z_1,z_2)\right] \ ,\label{eq:diseec2}
\end{eqnarray}
where the associated structure functions are written as
\begin{eqnarray}
        F_I^{h_1h_2}(x_B,z_1,z_2)&=&\int \frac{d\xi}{\xi}\frac{d\hat\xi}{\hat\xi^2}C_{ji}^I\left(\xi,\hat\xi,\mu\right) f_i\left(\frac{x_B}{\xi},\mu\right)\nonumber\\
        &&\times D_j^{h_1h_2}\left(\frac{z_1}{\hat\xi},\frac{z_2}{\hat\xi},\mu\right)\ ,
\end{eqnarray}
with the same hard coefficients as in SIDIS. Substituting this into Eq.~(\ref{eq:diseec0}), we derive the coefficients in Eq.~(\ref{eq:diseec1}): $C_{ji}^{\textrm{DIS},I}=\int d\hat\xi \hat\xi^2 C_{ji}^I(\xi,\hat\xi)$. Using the NLO coefficients $C_{ji}^I$ listed in the Appendix, we arrive at
\begin{eqnarray}
    &&C_{qq}^{\textrm{DIS},1,(1)}\nonumber\\
    &&~~~~=\delta (1-\xi)+\frac{\alpha_sC_F}{2\pi}\left[\frac{\hat P_{qq}(\xi)}{C_F}\ln\frac{Q^2}{\mu^2}+L_1(\xi)\right.\nonumber\\
    &&~~~~-L_2(\xi)-\frac{43}{12}\frac{1}{(1-\xi)_+}+\xi+\frac{19}{6}+\delta(1-\xi)\nonumber\\
    &&~~~~\left.\times\left(-\frac{11}{18}-\frac{\pi^2}{3}-\gamma_{qq}^{(3)}\ln\frac{Q^2}{\mu^2}\right)\right] \ ,
    \end{eqnarray}
    \begin{eqnarray}
    &&C_{gq}^{\textrm{DIS},1,(1)}\nonumber\\
    &&~~~~=\frac{\alpha_sC_F}{2\pi}\left[\frac{7}{12}\frac{1}{(1-\xi)_+}+\frac{1-2\xi}{6}+\delta(1-\xi)\right.\nonumber\\
    &&~~~~\left. \times \left(-{8\over 9}-\gamma_{gq}^{(3)}\ln\frac{Q^2}{\mu^2}\right)\right]\  ,
    \end{eqnarray}
    \begin{eqnarray}
    C_{qg}^{\textrm{DIS},1,(1)}&=&\frac{\alpha_s}{2\pi}\left[%
    \hat P_{qg}(\xi)\left(\ln\left(\frac{Q^2}{\mu^2}\frac{(1-\xi)}{\xi}\right)-\frac{5}{3}\right)\right.\nonumber\\
    &&\left. +\xi(1-\xi)\right] \ ,\\
    C_{qq}^{\textrm{DIS},L,(1)}&=&C_F\xi\ , \\
    C_{gq}^{\textrm{DIS},L,(1)}&=&\frac{1}{3}C_F\xi\ ,\\
    C_{qg}^{\textrm{DIS},L,(1)}&=&\frac{4}{3}\xi(1-\xi)\ .
\end{eqnarray}
The scale dependence in the above expressions results from the renormalization of the quark distribution $f_q(x,\mu)$. Additional scale dependence comes from the renormalization of the EEC jet function $\Gamma_q(\mu)$. However, this scale dependence is not complete, as we still need to take into account inhomogeneous contribution. 

The calculations for the inhomogeneous contribution are straightforward, and we find that the coefficients ${\cal D}_{i}^{I}$ in Eq.~(\ref{eq:diseec1}) can be obtained from the coefficients listed in the Appendix as ${\cal D}_i^I(\xi)=\int d\hat\xi \hat\xi(1-\hat\xi) \sum_j C_{ji}^I(\xi,\hat\xi)$. Applying this, we have for their ${\cal O}(\alpha_s)$ terms:
\begin{eqnarray}
    {\cal D}_q^{1,(1)}&=&\frac{\alpha_s}{2\pi}\left\{\delta(1-\xi)\left[-3+\left(\gamma_{qq}^{(3)}+\gamma_{gq}^{(3)}\right)\ln\frac{Q^2}{\mu^2}\right]\right.\nonumber\\
    &&\left. +\frac{3}{2}\frac{1}{(1-\xi)_+}-\frac{2\xi+1}{3}\right]\ , \\
    {\cal D}_g^{1,(1)}&=&\frac{\alpha_s}{2\pi}\frac{1}{3}\left(\xi^2+(1-\xi)^2\right) \ ,\\
    {\cal D}_q^{L,(1)}&=&\frac{2}{3}C_F\xi\ ,\\
    {\cal D}
    _g^{L,(1)}&=&\frac{4}{3}\xi(1-\xi)\ .
\end{eqnarray}
We have cross-checked the above results by directly computing them from the partonic cross sections with two individual fragmenation functions in the final state; 
see also the discussions in the next subsection.  

\begin{figure}[htbp]
  \begin{center}
\includegraphics[width=0.48\textwidth]{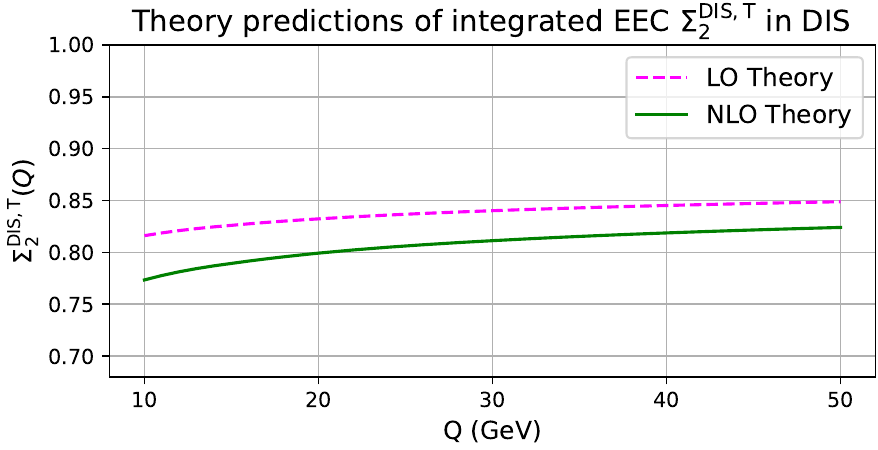} 
\caption{Integrated EEC in DIS, $\Sigma_2^{{\rm DIS},T}$, as function of $Q$ for kinematics at the future EIC with $x_B=0.1$.}
  \label{fig:integratedEEC_DIS}
 \end{center}
  \vspace{-5.ex}
\end{figure}

In Fig.~\ref{fig:integratedEEC_DIS}, we show the results of $\Sigma_2^{{\rm DIS},T}$ for the typical kinematics at the future EIC. Again, asymptotically as $Q\to \infty$, $\Gamma_q\sim \Gamma_g\to 1$, and  $\Sigma_2^{\rm DIS}$ approaches unity. Similar to $e^+e^-$ annihilation, the size of the NLO corrections is rather modest, reducing the correlation.

\subsection{Small-Angle EECs in DIS}

The angular distribution of the EECs is frame-dependent. In principle, one can work in any frame, including the target rest frame, the lepton-proton c.m.s. frame, or any other frame. In the following, we perform our calculations in the Breit frame where the momentum of the incoming photon only has a spatial component.

\begin{figure}[htbp]
  \begin{center}
   \includegraphics[scale=0.80]{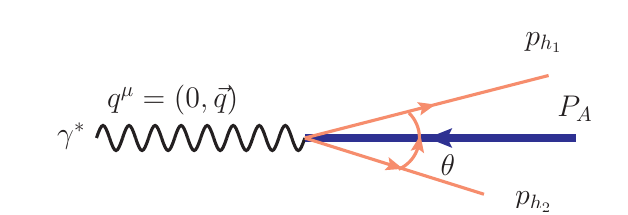} 
\caption{EEC in DIS defined in the Breit frame. The virtual photon moves along the $+\hat z$ direction. The two energy flows define the angle $\theta$.}
  \label{fig:eec_dis}
 \end{center}
  \vspace{-5.ex}
\end{figure}
As indicated in Fig.~\ref{fig:eec_dis}, the outgoing final-state hadrons will mainly emerge in the direction of the incoming photon. The angular distribution can be formulated accordingly,
\begin{eqnarray}
    \frac{d\Sigma_2^{\rm DIS}}{d\zeta}&=&\frac{1}{\sigma_{\textrm{tot}}}\sum_{h_1h_2}\int d\sigma \frac{p_{h_1}\cdot p_Np_{h_2}\cdot p_N}{(q\cdot p_N)^2}\nonumber\\
    &&~~~~~~~~~~~~\times \delta\left(\zeta-\frac{1-\cos(\theta)}{2}\right)\ ,\label{eq:zetadefsidis} 
\end{eqnarray}
where $\theta$ is the angle between the two hadrons $h_1$ and $h_2$ in the Breit frame; see Fig.~\ref{fig:eec_dis}.

Similar to the angular distribution in $e^+e^-$ annihilation, we can also compute the $\zeta$-distribution of the EEC in DIS in perturbative QCD, as
long as $\zeta\neq 0$. Because of the parton distributions involved in the DIS process, this is not an IR safe observable. Nevertheless, it can 
be derived perturbatively from $2\to 2$ partonic processes.

Let us take the $\gamma^* q\to qg$ channel as an example. In the Breit frame, the incoming photon's momentum can be written as $q^\mu=(0,0,0,Q)$, where $q^2=-Q^2$. Therefore, we have we have $q^+=-q^-=Q/\sqrt{2}$ for the light-cone components. The incoming quark has momentum $p_1=xp_N$ along the $-\hat z$ direction, and we have $p_1^+=0$, $p_1^-=Q/(\xi\sqrt{2})$, where $\xi=x_B/x$. Neither $q$ or $p_1$ has transverse-momentum components. On the other hand, the momenta of the final-state quark ($k_1$) and gluon ($k_2$) can be parameterized as
\begin{eqnarray}
    k_1&=&\hat \xi_1 q^++\frac{k_{1\perp}^2}{2\hat\xi_1 q\cdot p_1}p_1+k_{1\perp} \ ,\\
    k_2&=&\hat \xi_2 q^++\frac{k_{2\perp}^2}{2\hat\xi_2 q\cdot p_1}p_1+k_{2\perp} \ . 
\end{eqnarray}
At this order, we have $\hat \xi_1+\hat\xi_2=1$ and $\vec{k}_{1\perp}+\vec{k}_{2\perp}=0$ in the Breit frame. In addition, we have 
\begin{equation}
    k_{1\perp}^2=k_{2\perp}^2=\frac{Q^2(1-\xi)}{\xi}\hat\xi_1\hat\xi_2 \ .
\end{equation}
The invariant mass squared of two final-state partons is
\begin{equation}
    \hat s=(k_1+k_2)^2=\frac{Q^2(1-\xi)}{\xi}\ .\label{eq:disshat} 
\end{equation}
When the quark and gluon become collinear to each other, $\xi$ goes to 1 and the invariant mass vanishes. In the Breit frame, we can also compute the invariant mass from the energies of the two final-state partons. From momentum and energy conservation, we find
\begin{eqnarray}
   E_1&=&\frac{1}{\sqrt{2}}\left(k_1^++k_1^-\right)=\frac{k_{1\perp}^2+\hat \xi_1^2Q^2}{2\hat\xi_1Q} \ ,\\
   E_2&=&\frac{1}{\sqrt{2}}\left(k_2^++k_2^-\right)=\frac{k_{2\perp}^2+\hat \xi_2^2Q^2}{2\hat\xi_2Q} \ .
\end{eqnarray}
From this, $\hat s$ can be written as
\begin{eqnarray}
    \hat s&=&2E_1E_2(1-\cos(\theta_{12}))\nonumber\\
    &=&\frac{1-\cos(\theta_{12})}{2}\frac{(k_{1\perp}^2+\hat \xi_1^2Q^2)(k_{2\perp}^2+\hat \xi_2^2Q^2)}{\hat\xi_1\hat\xi_2Q^2}  \ ,\;\;
\end{eqnarray}
where $\theta_{12}$ is the angle between the two final-state particles in the Breit frame. Combining this equation with Eq.~(\ref{eq:disshat}), we find
\begin{equation}
    \zeta=\frac{1-\cos(\theta_{12})}{2}=\frac{\xi\bar\xi}{(\hat\xi_1\xi+\hat\xi_2\bar\xi)(\hat\xi_1\bar\xi+\hat\xi_2\xi)} \ , \label{eq:diszeta}
\end{equation}
where $\bar\xi=1-\xi$.

Having worked out all kinematics, we are ready to present the EEC differential in $\zeta$. The starting point is the cross section for single-inclusive hadron production in DIS (SIDIS). For example, for the SIDIS differential cross section associated with the $F_I$ structure function, we have 
\begin{equation}
    \frac{d\sigma^I}{dx_Bdydz_{h_1}}=\sigma_0^I\frac{\alpha_s}{2\pi}C_F\int\frac{d\xi}{\xi}\frac{d\hat\xi_1}{\hat\xi_1}f_q(x)D_q^{h_1}(z_1) \hat\sigma_I(\xi,\hat\xi_1) \ ,
\end{equation}
for the $\gamma^*q\to qg$ channel, where $x=x_B/\xi$ and $z_1=z_{h_1}/\hat\xi_1$. Here, $\sigma_0^I$ represents the leading order cross section normalization and $\hat\sigma_I$ the partonic cross section contribution for $\gamma^*q\to qg$. We can extend this expression to the cross section for di-hadron production, where two hadrons come from individual fragmentations of two partons:
\begin{eqnarray}
        &&\frac{d\sigma^I}{dx_Bdydz_{h_1}dz_{h_2}}\nonumber\\
        &&~~~=\sigma_0^I\frac{\alpha_s}{2\pi}C_F\int\frac{d\xi}{\xi}\frac{d\hat\xi_1}{\hat\xi_1}\frac{d\hat\xi_2}{\hat\xi_2}f_q(x)\hat\sigma_I(\xi,\hat\xi_1) \nonumber\\
        &&~~~~~\times D_q^{h_1}(z_1) D_g^{h_2}(z_2)\delta(\hat\xi_1+\hat\xi_2-1)\ ,
\end{eqnarray}
where $z_2=z_{h_2}/\hat\xi_2$. The inhomogeneous part of the EEC will be obtained from this by integrating out $z_{h_1}$ and $z_{h_2}$ with a weight factor  $z_{h_1}z_{h_2}$,
\begin{eqnarray}
    \Sigma_2^{\textrm{DIS,inhom.}}&=&\frac{1}{\sigma_{\textrm{tot}}}\int dz_{h_1}dz_{h_2} (z_{h_1}z_{h_2})\frac{d\sigma^I}{dx_Bdydz_{h_1}dz_{h_2}}\nonumber\\
&=&\frac{2}{F_I(x_B)}\frac{\alpha_s}{2\pi}C_F\int\frac{d\xi{d\hat\xi_1}}{\xi}{\hat\xi_1(1-\hat\xi_1)}f_q(x)\nonumber\\ 
&&\times \hat\sigma_I(\xi,\hat\xi_1) \ ,
\label{eq:S2SIDIS}
\end{eqnarray}
where the factor $2$ represents the two configurations for fragmentation of the final-state partons, for instance, $qg$ and $gq$. Inserting the partonic differential cross section and fully carrying out the integrals, we recover the inhomogeneous contribution to the integrated EEC presented in the previous subsection. More precisely, in dimensional regularization, the unpolarized partonic differential cross section for $\gamma^*q\to qg$ corresponding to the $F_1$ structure function can be written as
\begin{eqnarray}
    \hat\sigma_1&=&\frac{1}{\bar\xi\hat\xi_2}\left\{1+\xi^2\hat\xi_1^2+\bar\xi^2\hat\xi_2^2-2\epsilon\left(1-\xi\bar\xi-\hat\xi_1\hat\xi_2\right)\right.\nonumber\\
    &&\left.~~~~+{\cal O}(\epsilon^2)\right\} \ .
\end{eqnarray}
Upon insertion into Eq.~(\ref{eq:S2SIDIS}), 
the integral over $\hat\xi_1$ leads to a $1/(1-\xi)$ behavior. Using the expansion
\begin{equation}
    \frac{1}{(1-\xi)^{1+\epsilon}}=-\frac{1}{\epsilon}\delta(1-\xi)+\frac{1}{(1-\xi)_+} +\cdots \ ,
\end{equation}
we can separate out the IR divergence which is absorbed into the renormalization of the EEC jet function. The results for all other channels can be derived accordingly.

We now turn to the angular distribution. For convenience, we follow a similar notation as for the EEC in $e^+e^-$ annihilation, using the variable $\zeta$ as defined in Eq.~(\ref{eq:zetadefsidis}) above. Since $\zeta$ is completely determined by $\xi$ and $\hat\xi_1$ through Eq.~(\ref{eq:diszeta}), the distribution in $\zeta$ can be computed as
\begin{eqnarray}
    \frac{d\Sigma_2^{\textrm{DIS,inhom.}}}{d\zeta}&=&\frac{2}{F_I(x_B)}\frac{\alpha_s}{2\pi}C_F\int\frac{d\xi{d\hat\xi_1}}{\xi}{\hat\xi_1(1-\hat\xi_1)}\nonumber\\
    &&\times f_q(x)\hat\sigma_I(\xi,\hat\xi_1)\delta(\zeta-\zeta_\xi) \ \label{eq:disfixedorder} \ ,
\end{eqnarray}
where $\zeta_\xi$ is given by Eq.~(\ref{eq:diszeta}),
\begin{equation}
    \zeta_\xi\equiv\frac{\xi\bar\xi}{(\hat\xi_1\xi+\hat\xi_2\bar\xi)(\hat\xi_1\bar\xi+\hat\xi_2\xi)}  \ .
\end{equation}
For any finite $\zeta$, the above gives a finite result. However, when $\zeta$ goes to zero, a collinear divergence appears. Again, we can apply dimensional regularization, to obtain
\begin{equation}
     \left.\frac{d\Sigma_2^{\textrm{DIS,inhom.}}}{d\zeta}\right|_{\zeta\to 0}=
     \frac{\alpha_s}{2\pi}\left(\gamma_{qq}^{(3)}+\gamma_{gq}^{(3)}\right)\frac{1}{\zeta^{1+\epsilon}} \ ,\label{eq:disasymptotic}
\end{equation}
where the factor $\zeta^{-\epsilon}$ comes from the azimuthal-angle integral over the final-state particles. 
The collinear divergence at $\zeta\to 0$ needs to be subtracted by renormalization of the unintegrated EEC jet function at small angle. 

At the leading order, the final-state quark moves along the $+\hat z$ direction with energy and momentum $Q/2$. This is very similar to the case of $e^+e^-$ annihilation discussed in the previous section. The leading-order result can again be written as (cf. Eq.~(\ref{eq:NearEEC}))
\begin{equation}
    \frac{d\Sigma_2^{\rm DIS}}{d\zeta}=\pi Q^2\Gamma_q^{(0)}(  Q,q_T)\ ,
\end{equation}
where $\overline Q=Q/2$ and $q_T^2=\zeta Q^2$. The NLO  calculations are similar to those for the inclusive di-hadron differential cross section, while the factorization follows that for the EEC in $e^+e^-$ annihilation, including both homogeneous and inhomogeneous contributions. The former takes the following form:
\begin{eqnarray}
    \frac{d\Sigma_2^{\rm DIS,hom.}}{d\zeta}&=&\frac{\pi Q^2}{\sigma_{\rm incl.}}\frac{2\pi\alpha^2}{Q^2}\left[\frac{1+(1-y)^2}{y}2F_1^{\Sigma_2}(x_B,\zeta)\nonumber\right.\\
    &&~~~\left.+\frac{2(1-y)}{y}F_L^{\Sigma_2}(x_B,\zeta)\right] \ ,
\end{eqnarray}
where, using the same factorization as for semi-inclusive hadron production, 
\begin{eqnarray}
        F_I^{\Sigma_2}(x_B,\zeta)&=&\int \frac{d\xi}{\xi}{d\hat\xi_1}{\hat\xi_1^2\hat\xi^{\prime 2}}f_i\left(\frac{x_B}{\xi},\mu\right)\Gamma_j(\mu,q_T)\nonumber\\
        &&\times C_{ji}^I(x,\hat\xi_1,\mu) \ ,
\end{eqnarray}
again with $I=L,T$. Here, $q_T^2=\zeta (\hat\xi'Q)^2$ with $\hat\xi'=(1+2\xi\hat\xi_1-\xi-\hat\xi_1)/\xi$ representing the ratio of energy of the fragmenting parton over $\overline Q$ in the Breit frame. We emphasize that this ratio is frame dependent, as are $\theta_{12}$ and $\zeta$. 
In the limit of $\xi\to 1$ or $\hat\xi_1\to 1$, we have $\hat\xi_1'\to \hat\xi_1$. The above relation for $\hat{\xi}'$ only holds at the first order which we are considering here. At higher orders, it will become more complicated. 

We can also work out the factorization obtained for the Fourier transform to $b_T$-space:
\begin{eqnarray}
        \widetilde{F}_I^{\Sigma_2}(x_B,b_T)&=&\int \frac{dx}{x}{d\hat\xi_1}{\hat\xi_1^2}f_i\left(\frac{x_B}{x},\mu\right)\widetilde{\Gamma}_j\left(\mu,{b_T\over \hat\xi'}\right)\nonumber\\
        &&\times C_{ji}^I(x,\hat\xi_1,\mu) \ .
\end{eqnarray}
Similar to Sec.~\ref{sec:nearEEC} this expression can again be used to carry out
a resummation of $b_T$-logarithms in, for instance, the LLA.

\begin{figure}[htbp]
  \begin{center}
   \includegraphics[width=0.48\textwidth]{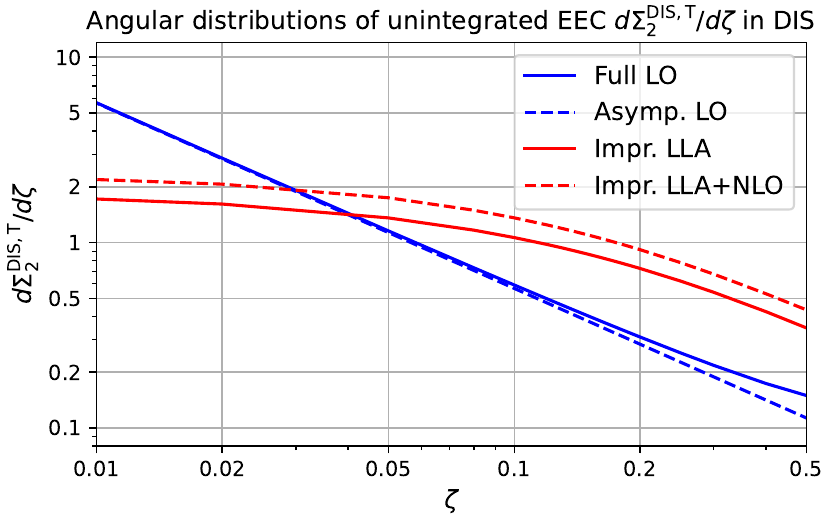} 
\caption{Angular distribution of the EEC in DIS in the Breit frame. We have chosen $x_B=0.1$ and $Q=10$~GeV.}
  \label{fig:EEC_DIS_10}
 \end{center}
  \vspace{-5.ex}
\end{figure}

Figure~\ref{fig:EEC_DIS_10} shows the angular distribution of the EEC in DIS in the Breit frame, using kinematics relevant at the future EIC. Here ``full LO'' stands for the result given by the inhomogeneous contribution, Eq.~(\ref{eq:disfixedorder}), whereas the ``asymptotic LO'' is given by Eq.~(\ref{eq:disasymptotic}) with $\epsilon=0$. Clearly, the fixed-order result is reasonably reproduced by the asymptotic one. We observe that the LLA resummation effects, shown here as the ratio of $F_1^{\Sigma_2}(x_B,\zeta)/F_1(x_B)$, are quite pronounced, leading to the expected decrease of the EEC at lower $\zeta$, and to an increase toward higher $\zeta$. Again, the NLO corrections are about $20\%$ when $\zeta\to 0$, which can be estimated by comparing $\int_0^1 d\hat\xi_1 \hat\xi_1^4 \left[C_{qq}^1(\xi,\hat\xi_1)+C_{gq}^1(\xi,\hat\xi_1)\right]$ to $\int_0^1 d\hat\xi_1  C_{qq}^1(\xi,\hat\xi_1)$, similar to the case of the EEC in $e^+e^-$ annihilation in the previous section. Clearly, the EECs in DIS at the future EIC will be very sensitive to the EEC jet functions, which will provide important constraints on the non-perturbative features of these functions. 

Meanwhile, we note that the existing data from the HERA collider would be interesting to look into as well. A preliminary analysis of energy correlators in DIS processes from HERA experiments has been reported recently~\cite{Miguel:HERA}, although in a somewhat different setting from what we discussed above. It would be very interesting if our proposed observables given by Eqs.~(\ref{eq:eecdisdef0}) and (\ref{eq:zetadefsidis}) (see also the illustration of the EECs in DIS in Fig.~\ref{fig:eec_dis}), could be studied for archival HERA data. Among other things, this would be important to test the universality of the EEC jet functions. Such analyses would also provide important guidelines for future measurements at the EIC.

\section{Conclusion}
We have investigated energy–energy correlators (EECs) in electron–positron annihilation and deep-inelastic scattering. Focusing on perturbative corrections to the correlators, we have derived NLO expressions for them. Although the EECs are infrared safe at fixed angles, integrating over the angle introduces infrared sensitivity related to di-hadron fragmentation functions. To handle the collinear limit, we have introduced EEC jet functions constructed from these fragmentation functions, which turn out to be universal across the two different processes we have considered. The scale dependence of the (integrated) jet functions is surprisingly mild, which is primarily due to momentum-conservation properties of the underlying fragmentation functions. 

Our analysis includes explicit one-loop calculations and the derivation of evolution equations in transverse-coordinate $b_T$ space, enabling resummation of logarithms in $b_T$ for the unintegrated EEC jet functions. The resummation for the unintegrated EEC jet functions can be derived analytically for two distinct kinematic regions. In the perturbative regime ($1/b_T\sim q_T\gg\Lambda_{\rm QCD}$), we can apply the leading logarithmic approximation, and the scaling behavior is controlled by the relevant anomalous dimension $\gamma_{ij}^{(3)}$. However, in the non-perturbative post-confinement region ($q_T\sim 0$), the scaling behavior is controlled by a different anomalous dimension, $\gamma_{ij}^{(5)}$. Observing this difference, we have proposed a matching between the two regions, which smoothly connects the perturbative and nonperturbative regimes. 

With only one parameter determining the matching point, our proposed scheme provides a very good description of PYTHIA simulations for the EECs in $e^+e^-$ annihilation over a wide range of energies. Sizable NLO corrections are also found in these comparisons, calling for future computations of even higher-order corrections. Meanwhile, it is also important to solve the full evolution equation. Further, more detailed phenomenological
studies will be needed in order to assess the opportunities and potential that the future EIC presents for the study of EECs.

\section{Acknowledgment}

We thank Daniel de Florian and Felix Ringer for helpful discussions and comments. This work is supported by the Office of Science of the U.S. Department of Energy under Contract No. DE-AC02-05CH11231, and by the U.S. Department of Energy, Office of Science, Office of Nuclear Physics, within the framework of the Quark Gluon Tomography (QGT) Topical Theory Collaboration. This work has been supported by Deutsche Forschungsgemeinschaft (DFG) through the Research Unit FOR 2926 (Project No. 409651613).

\bibliographystyle{apsrev4-1}
\bibliography{refs.bib}

\newpage

\appendix

\section{Hard Coefficients at One-loop Order}

For convenience, we list the first-order coefficients computed in Ref.~\cite{deFlorian:2003cg} for the $e^+e^-$ annihilation process,
\begin{eqnarray}
C_q^{e^+e^-}(x) & = & %
C_F \left[(1+x^2)\left (\frac{\ln\left(1-x\right)}{1-x}\right)_+%
- \frac{3}{2}\frac{1}{(1-x)_+}\right.\nonumber\\
&&%
+ 2 \left ( \frac {1+x^2}{1-x}\right )\ln x -  \frac{3}{2}x + \frac{5}{2}\nonumber\\
&&\left.+   %
\left (\frac{2\pi^2}{3} - \frac{9}{2}\right )\delta (1-x) \right],\\
 C^{e^+e^-}_g(x)  & = & \hat{P}_{gq}(x) \ln(x^2(1-x) ), \label{fg}
\end{eqnarray}
where the splitting kernels are given by
\begin{eqnarray}
    \hat P_{qq}(x)&=&C_F\left(\frac{1+x^2}{1-x}\right)_+\ , \\
    \hat P_{gq}(x)&=&C_F\frac{1+(1-x)^2}{x} \ , \\
    \hat P_{gg}(x)&=&2C_A\frac{(1-x+x^2)^2}{x(1-x)_+}+\frac{\beta_0}{2}\delta(1-x)\ ,\\
    \hat P_{qg}(x)&=&\frac{1}{2}
    \left(x^2+(1-x)^2\right)\  ,
\end{eqnarray}
with $\beta_0=\frac{11}{3}C_A-\frac{2}{3}n_f$.

For the EECs in DIS, we have applied the hard coefficients computed for SIDIS~\cite{Furmanski:1981cw,deFlorian:1997zj},
whose NLO contributions read
\begin{eqnarray}
 &&   C_{qq}^{1,(1)}\nonumber\\
 &=& C_F\left\{-8\delta(1-\xi)\delta(1-\hat\xi)+\delta(1-\xi)\left[\frac{\hat P_{qq}(\hat\xi)}{C_F}\ln\frac{Q^2}{\mu^2}\right.\right.\nonumber\\
    &&\left.+L_1(\hat\xi)+L_2(\hat\xi)+(1-\hat\xi)\right]+\delta(1-\hat\xi)\left[\frac{\hat P_{qq}(\xi)}{C_F}\ln\frac{Q^2}{\mu^2}\right.\nonumber\\
    &&\left.+L_1(\xi)-L_2(\xi)+(1-\xi)\right]+2\frac{1}{(1-\xi)_+}\frac{1}{(1-\hat\xi)_+}\nonumber\\
    &&\left.-\frac{1+\hat\xi}{(1-\xi)_+}-\frac{1+\xi}{(1-\hat\xi)_+}+2(1+\xi\hat\xi)\right\} \ , \\
   && C_{gq}^{1,(1)}\nonumber\\
   &=& C_F\left\{\frac{\hat P_{gq}(\hat\xi)}{C_F}\left[\delta(1-\xi)\ln\left(\frac{Q^2}{\mu^2}\hat\xi(1-\hat\xi)\right)+\frac{1}{(1-\xi)_+}\right]\right.\nonumber \\
    &&\left.+\hat\xi\delta(1-\xi)+2(1+\xi-\xi\hat\xi)-\frac{1+\xi}{\hat\xi}\right\} \ , 
\end{eqnarray}
\begin{eqnarray}
    C_{qg}^{1,(1)}&=& %
    \delta(1-\hat\xi)\left[\hat P_{qg}(\xi)\ln\left(\frac{Q^2}{\mu^2}\frac{1-\xi}{\xi}\right)+\xi(1-\xi)\right]%
    \nonumber \\
    &&\left.+\hat P_{qg}(\xi)\left[\frac{1}{(1-\hat\xi)_+}+\frac{1}{\hat \xi}-2\right]\right\}\ ,   \\
    C_{qq}^{L,(1)}&=&4C_F\xi\hat\xi  \ ,   \\
    C_{gq}^{L,(1)}&=&   4C_F\xi(1-\hat\xi)\ ,  \\
    C_{qg}^{L,(1)}&=&    4\xi(1-\xi) \ .
\end{eqnarray}

\section{Derivation of Eq.~(\ref{eq:eeinhomo})}

We start with the two-parton differential cross section in $e^+e^-$ annihilation,
\begin{eqnarray}
    \frac{d\sigma}{dx_1dx_2}&=&\sigma_0\frac{\alpha_s}{2\pi}\frac{C_F}{\Gamma(2-\epsilon)}\left(\frac{Q^2}{4\pi\mu^2}\frac{(1-z^2)x_1^2x_2^2}{4}\right)^{-\epsilon}\nonumber\\
    &&\times F(x_1,x_2) \ ,    
\end{eqnarray}
where $x_1$ and $x_2$ are the energy fractions carried by the quark and the antiquark in the final state, respectively, and $z=\cos(\theta_{12})$ with $
\theta_{12}$ the angle between them. At this order, because of energy conservation, we have $x_1+x_2+x_3=2$ and $z=1+2(1-x_1-x_2)/x_1x_2$. The function $F(x_1,x_2)$ is 
given by
\begin{eqnarray}
    F(x_1,x_2)&=&(1-\epsilon)^2\frac{x_1^2+x_2^2}{x_1x_2}\nonumber\\
    &&-2\epsilon(1-\epsilon)\frac{2-2x_1-2x_2+x_1x_2}{(1-x_1)(1-x_2)} \ .    \;\;
\end{eqnarray}
The inhomogeneous contribution to di-hadron production can be expressed in terms of the above partonic cross section as
\begin{equation}
    \frac{d\sigma^{\mathrm{inhom.}}}{dz_1dz_2}=\int_{x_1+x_2>1} \frac{dx_1}{x_1}\frac{dx_2}{x_2}D_i(\frac{z_1}{x_1})D_j(\frac{z_2}{x_2})\frac{d\sigma^{ij}}{dx_1dx_2} \ ,
\end{equation}
where $ij$ run over $q\bar q$, $qg$, $\bar qg$ configurations. Now, for the integrated EEC,
\begin{eqnarray}
    \Sigma_2^{\textrm{inhom.}}&=&\frac{1}{\sigma}\int dz_1 dz_2 z_1 z_2\frac{d\sigma}{dz_1dz_2}\nonumber\\
    &=&\frac{1}{\sigma}\int dx_1dx_2 x_1x_2\sum_{ij}\frac{d\sigma^{ij}}{dx_1dx_2} \ .
\end{eqnarray}
Inserting the above differential cross sections, we have
\begin{eqnarray}
    \Sigma_2^{\textrm{inhom.}}&=&\frac{\sigma_0}{\sigma}\frac{\alpha_s}{2\pi}\frac{C_F}{\Gamma(2-\epsilon)}\left(\frac{Q^2}{4\pi\mu^2}\right)^{-\epsilon}\int dx_1 dx_2 (x_1x_2)\nonumber\\
    \nonumber\\
    &&\times \left(\frac{(1-z^2)x_1^2x_2^2}{4}\right)^{-\epsilon}\left[F(x_1,2-x_1-x_2)\right.\nonumber\\
    &&\left.+F(x_1,x_2)+F(2-x_1-x_2,x_2)\right] \nonumber\\
    &&+\frac{\sigma_0}{\sigma}\frac{\alpha_s}{2\pi}C_FN_\epsilon\left(-\frac{2}{\epsilon^2}-\frac{3}{\epsilon}-8+\pi^2\right)
\ ,
\end{eqnarray}
where the last term is the virtual contribution. The above integral can be carried out by writing the variable $x_2=1-vx_1$. With that, the condition $x_1+x_2>1$ 
leads to $0<x_1<1$, $0<v<1$. We can also switch to $\zeta=(1-z)/2$; where then $0<\zeta<1$. The angular distribution for the EEC can be calculated as differential
with respect to $\zeta$:
\begin{eqnarray}
    \frac{d\Sigma}{d\zeta}&=&\int dx_1\frac{x_1(1-x_1)}{(1-x_1\zeta)^2}x_1x_2
    \frac{\alpha_s}{2\pi}C_F\left[\frac{x_1^2+x_2^2}{(1-x_1)(1-x_2)}\right.\nonumber\\
    &+&\left. (x_2\to 2-x_1-x_2)+(x_1\to 2-x_1-x_2)\right] \ .
\end{eqnarray}
This leads to the LO result for the EEC angular distribution as function of $\zeta$,
\begin{eqnarray}
    \frac{d\Sigma}{d\zeta}&=&\frac{\alpha_s}{2\pi}C_F\frac{3-2\zeta}{\zeta^5(1-\zeta)}\left[3\zeta(2-3\zeta)+2(2\zeta^2-6\zeta+3)\right.\nonumber\\
    &&\left.\times \ln(1-\zeta)\right] \ .
\end{eqnarray}
As it stands, this result cannot be integrated over $\zeta$ because of the singular behavior at $\zeta\to 1$. To obtain
a properly regularized expression, we use dimensional regularization. We then have, in $d=4-2\epsilon$ dimensions,
\begin{eqnarray}
    \frac{d\Sigma}{d\zeta}&=&\frac{\alpha_s}{2\pi}C_FN_\epsilon\left\{\frac{3}{2}\frac{1}{\zeta^{1+\epsilon}}-2\left(\frac{\ln(1-\zeta)+3/2}{(1-\zeta)}\right)_+\right.\nonumber\\
    &&\left.+\frac{37}{6}\frac{\epsilon}{\zeta^{1+\epsilon}}+\delta(1-\zeta)\left(-4-\frac{\pi^2}{3}\right)+\Sigma_f\right\}\ ,\nonumber\\
\end{eqnarray}
where $N_\epsilon=\frac{\Gamma(1+\epsilon)\Gamma^2(1-\epsilon)}{\Gamma(1-2\epsilon)}(\frac{Q^2}{4\pi\mu^2})^{-\epsilon}$ and
\begin{eqnarray}
    \Sigma_f&=&\frac{1}{2}\left\{\frac{36}{\zeta^5}\left(\zeta+\frac{\zeta^2}{2}+\frac{\zeta^3}{3}+\frac{\zeta^4}{4}+\ln(1-\zeta)\right)\right.\nonumber\\
    &&
    -\frac{60}{\zeta^4}\left(\zeta+\frac{\zeta^2}{2}+\frac{\zeta^3}{3}+\ln(1-\zeta)\right)\nonumber\\
    &&
    +\frac{12}{\zeta^3}\left(\zeta+\frac{\zeta^2}{2}+\ln(1-\zeta)\right)\nonumber\\
    &&\left.-\frac{4}{\zeta^2}\left(\zeta+\ln(1-\zeta)\right)-\frac{4\ln(1-\zeta)}{\zeta}\right\}\ .
\end{eqnarray}

\end{document}